\documentstyle[preprint,aps,amssymb,emlines,bezier,epsfig]{revtex}
\tighten
\begin{document}
%%%%%%%%%%%%%%%%
\title{Three--Body Configuration Space Calculations with
       Hard Core Potentials}
\author{E. A. Kolganova}
\address{ Laboratory of Computing Techniques and Automation\\
Joint Institute  for Nuclear Research, Dubna, 141980, Russia}
\author{ A. K. Motovilov\thanks{On leave of absence from
         the Laboratory of Theoretical Physics,
    Joint Institute  for Nuclear Research, Dubna, 141980, Russia},
	S. A. Sofianos}
\address{Physics Department, University of South Africa,  P.O.Box
	  392, Pretoria 0001, South Africa}
%%%%%%%%%%%%%%%%%%%%%%%%%%%%
\date{February 26, 1998} %%
%%%%%%%%%%%%%%%%%%%%%%%%
%%%%%%%%%%%%%%%%%%%%%%%%
%%%%%%%%%%%%%%%%%%%%%%%%
\maketitle
%%%%%%%%%%%%%%%%%%%%% ABSTRACT %%%%%%%%%%%%%%%%%%%%%%%%%%%%%%%%%%%
\begin{abstract}
%%%%%%%%%%%%%%%
We present a mathematically rigorous  method suitable  for
solving three-body bound state and scattering problems when the
inter--particle interaction is of a hard--core nature. The
proposed method is a variant of the Boundary Condition Model and
it has been employed to calculate the binding energies for a
system consisting of three $^4$He atoms. Two realistic
\mbox{He--He} interactions of Aziz and collaborators, have been
used for this purpose.  The results obtained compare favorably
with those previously obtained by other methods. We further
used the model to calculate, for the first time, the ultra-low
energy scattering phase shifts.  This  study revealed that
our method is ideally suited for three-body molecular
calculations  where the practically  hard--core  of the
inter--atomic potential gives rise  to strong  numerical
inaccuracies that make  calculations for these molecules
cumbersome.\\\\

\noindent LANL E-print physics/9612012. 

\noindent
Published in J. Phys. B., 1998, v.~31, Nr.~6, pp.~1279--1302

%{PACS numbers: 02.60.Nm, 21.45.+v, 34.40.-m, 36.40.+d}
\end{abstract}
%%%%%%%%%%%%%%%%%%%%%%%%%%%%%%%%%%%%%%%%%%%%%%%%%%%%%%%%%%%%%%%%%%
\section{Introduction}
\label{SIntro}
%%%%%%%%%%%%%%%%%%%%%%%%%%%%%%%%%%%%%%%%%%%%%%%%%%%%%%%%%%%%%%%%%%%%%
The Boundary Condition Model (BCM) (see, for example,
Refs.~\cite{EfiSch,MMYa}) is of interest due to its simplicity
in describing the short-range component of the interaction
between particles.  In the BCM the interaction is specified by
boundary conditions imposed on the wave function when the
particles approach each other at a certain distance $r=c$. The
so-called hard core potentials represent a particular variant of
the BCM where one requires that the wave function vanishes at
$r=c$. Such a requirement is equivalent to an introduction of an
infinitely strong repulsion between particles at distances
$r\leq c$. The standard formalism for scattering
\cite{Faddeev63,MF} does not deal with  hard-core interactions
described by these boundary conditions. Therefore a derivation
of special equations  to handle this class of interactions is
desirable.

Replacement of the finite, for $r>0$, but often singular at
$r=0$, repulsive short-range part of the potential with a
hard-core interaction turns out to be a very effective way to
suppress inaccuracies related to a numerical approximation of
the Schr\"odinger operator at short distances. Although in
two--body applications these potentials are easy to handle, in
 three-body systems certain mathematical difficulties appear
 \cite{FaddeevMIFI}, which are absent when conventional
potentials are used.

To overcome these difficulties various approaches were
considered.  We shall recall here the two main ones related to
the Faddeev equations \cite{Faddeev63,MF}. In the first, a
certain limiting procedure is used where special potentials that
include only a finite repulsive core are constructed at a first
step. The parameters of these potentials are then  chosen so
that the final two-body wave function satisfies the desired
boundary conditions \cite{KimT,Belyaev,Brayshaw}. The
corresponding two-body $t$--matrices are subsequently
substituted into the Faddeev integral equations
\cite{KimT,Brayshaw} under an implicit assumption that the
latter are still valid. The resulting equations are considered
as a generalization of the Faddeev equations for the BCM. A
similar approach was also used in Refs.
\cite{EfiSch,VEfimov,KuKhar}.  A common feature of the reduced
equations in these  approaches is that  they are not of
a Fredholm type and that they have not a unique solution at
all energy values, including the complex ones.  To obtain a
unique solution one is forced to introduce auxiliary conditions
or relations \cite{EfiSch,KuKhar}.

In the second approach,  three--body integral equations of a
Fredholm type are derived in the BCM model without any limiting
procedure.  Instead, one uses the fact that the spectral problem
for the Schr\"odinger operator is an example of a
classical boundary--value problem for elliptic differential
equation in partial derivatives of the second order.  One of the
traditional methods to deal with such problems is the Potential
Theory \cite{Potential}.  An approach to the three--body problem
in the BCM which is based  on the Potential Theory was developed
in Refs. \cite{MerMot,MotovilovVLGU,MM-Kalinin,MotovilovThesis}
(see also \cite{MMYa} and \cite{EChAYa}). However, in contrast
to the boundary value problems for compact surfaces, the
initial three-body equations were not of Fredholm
type, similarly to the three--body Lippmann--Schwinger equation
in the case of the conventional soft--core potentials.  This is
due to the same reason, i.e, the noncompactness  of the support
of  the two-body interaction in the three-body configuration
space.  To overcome this problem,  these equations were
rearranged in Refs.
\cite{MerMot,MotovilovVLGU,MM-Kalinin,MotovilovThesis} using the
Faddeev method \cite{Faddeev63}.  The resultant equations  are
of Fredholm type and  suitable for use  in the three-particle
scattering problem.  These developments allowed the
reformulation of the Faddeev equations  for the  bound state and
scattering problems  in configuration space in terms of
boundary-value problems which  are suitable for  numerical
calculations.  This was demonstrated  in three-nucleon
bound-state  and scattering calculations
\cite{MotovilovVLGU,EChAYa}.

In this  work we shall present a hard core version of the BCM 
formalism \cite{MerMot,MotovilovVLGU,MM-Kalinin,MotovilovThesis} 
and apply it to the three-atomic $^4$He system.  Various methods 
have been  used in the past to study the ground state properties 
of $^4$He molecules.  We mention here the Variational Method 
(VM) \cite{McMillan,Schmid,VM3,VM4}, the Variational Monte Carlo 
method (VMC)  \cite{VMC1,VMC2}, the Green Function Monte Carlo 
method (GFMC) \cite{VMC2,GF2,GF3,GF4,Kalos,GF6}, the methods 
based on the Faddeev integral equations in momentum 
space~\cite{Nakai,Gloeckle}, the Faddeev differential equations 
in configuration space \cite{CGM}, and the hyperspherical 
methods~\cite{Levinger,Sof,EsryLinGreene}. 

The general atom-diatom collision problem has been addressed by 
various researchers in the field and we refer the interested reader to
the  review articles on this topic by Micha \cite{MichaRev81}
and Kuppermann~\cite{Kuppermann}. Collision dynamics at thermal energies
of the H$+$H$_2$ system  and the existence of resonances  
were discussed by Kuruoglu and Micha ~\cite{KuruogluMicha} using 
the Faddeev integral equations in momentum space. Finally, the
problem of existence  of the $^4$He $n$-mers and its relation 
to the Bose-Einstein condensation in He {\small II} was discussed 
in Refs.~\cite{GhassibChester,March}. From the experimental works we 
mention those of Refs.~\cite{ArgonExp,XenonExp,DimerExp,Science} where 
molecular clusters consisting of a small number of noble gas 
atoms were investigated.  
 
The interaction between bosons in such clusters is usually
described by central potentials having a very strong repulsive
cores~\cite{Huber78,Aziz79,Aziz87,Aziz91,Tang95}. In the
present work, we approximate the strong repulsion between the
Helium atoms at short distances by a hard core and solve the
corresponding boundary value problems for the Faddeev--type
differential equations. We shall show that the method gives
excellent results for the ground-state energy of the Helium
$^4$He trimer. It further allowed the  calculation of an excited
state interpreted in~\cite{Gloeckle,EsryLinGreene} as an Efimov
one~\cite{Efimov}. Moreover, we shall demonstrate that the
method is suitable for scattering calculations at ultra-low
energies below as well as above the breakup threshold.
Certain  results of our  work were presented in \cite{MSK-CPL}. 

Some comments on the notation used throughout the paper: The
$\sqrt{z}$,  $z\in{\Bbb C}$, stands for the main branch of the
function $z^{1/2}$. The $\hat{\bf a}$ denotes  the unit vector,
$\hat{\bf a}=\displaystyle\frac{\bf a}{|{\bf a}|},$ ${\bf
a}\in{{\Bbb R}}^n$, while $L_2(D)$ is the standard notation used for
the Hilbert space of the square integrable functions defined in
a domain $D$ of ${{\Bbb R}}^n$. The symbol $W_2^2(D)$ stands for the
space of those of the functions of $L_2(D)$ which have all
second partial derivatives as elements of $L_2(D)$.  Finally,
the notation $\overline{D}$ is used for closure of a set
$D\subset{{\Bbb R}}^n$.

This paper is organized as follows. In Sec.~II we overview the
three-body bound and scattering state formalism for the hard
core interactions and in Sec.~III, we describe its application
to a system of three identical bosons. Our numerical results for
the three-atomic $^4$He system are presented in Sect.~IV while
our conclusions are drawn in Sec.~V. A detailed description of
the numerical methods used is given in the Appendix.

%%%%%%%%%%%%%%%%%%%%%%%%%%%%%%%%%%%%%%%%%%%%%%%%%%%%%%%%%%%%%%%%%%%%%%%%
\section{Three-particle systems with hard core interactions}
%%%%%%%%%%%%%%%%%%%%%%%%%%%%%%%%%%%%%%%%%%%%%%%%%%%%%%%%%%%%%%%%%%%%%%%%
In describing the three-body system we use the standard  Jacobi
coordinates \cite{MF} ${\bf x}_{\alpha},{\bf y}_{\alpha}$,
$\alpha=1,2,3$, expressed in terms of the position vectors of
the particles ${\bf r}_i\in{{\Bbb R}}^3$ and their masses ${\rm m}_i$,
\begin{eqnarray}
\label{Jacobi}\nonumber
                {\bf x}_\alpha &=&
                \left[ \frac{2{\rm m}_\beta{\rm m}_\gamma}
    {{\rm m}_\beta + {\rm m}_\gamma} \right]^{1/2}
  ({\bf  r}_\beta - {\bf  r}_\gamma)\\&&\\
\nonumber
       {\bf  y}_\alpha & =&  \left[\frac
      {2{\rm m}_\alpha({\rm m}_\beta + {\rm m}_\gamma)}
       {{\rm m}_\alpha + {\rm m}_\beta + {\rm m}_\gamma}\right]^{1/2}
     \left( {\bf  r}_\alpha -\frac{{\rm m}_\beta{\bf  r}_\beta +
   {\rm m}_\gamma{\bf r}_\gamma} {{\rm m}_\beta + {\rm m}_\gamma}\right)
\end{eqnarray}
where $(\alpha,\beta,\gamma)$ stands for a cyclic permutation of
the indices $(1,2,3)$. The coordinates ${\bf x}_\alpha,{\bf
y}_\alpha$ fix the six-dimensional vector $X\equiv ({\bf
 x}_{\alpha},{\bf y}_{\alpha})\in{{\Bbb R}}^6$.  The vectors ${\bf
 x}_{\beta},{\bf y}_{\beta}$ corresponding to the same point $X$
as the pair ${\bf x}_{\alpha},{\bf y}_{\alpha}$ are obtained
using the transformations
$$
 		{\bf x}_{\beta}={\sf c}_{\beta\alpha} {\bf x}_{\alpha}+
     		{\sf s}_{\beta\alpha} {\bf y}_{\alpha}
\qquad
      		{\bf y}_{\beta}=-{\sf s}_{\beta\alpha} {\bf x}_{\alpha} +
      		{\sf c}_{\beta\alpha}{\bf y}_{\alpha}
$$
where the coefficients  ${\sf c}_{\beta\alpha}$ and ${\sf s}_{\beta\alpha}$
 fulfill the conditions
        $ -1<{\sf c}_{\beta\alpha} < +1$
and
        	${\sf s}_{\beta\alpha}^2=1-{\sf c}_{\beta\alpha}^2 $
with
       		${\sf c}_{\alpha\beta}={\sf c}_{\beta\alpha}$,
                ${\sf s}_{\alpha\beta}=-{\sf s}_{\beta\alpha}$,
$\beta\neq\alpha$
and depend only on the particle masses \cite{MF}. For equal masses
$ {\sf c}_{\beta\alpha}=-1/2$.

The configuration space $\Omega$ of the three-body system in the
hard-core model represents only a part of the space ${{\Bbb R}}^{6}$
external, $|{\bf x}_{\alpha}|> c_{\alpha}$, with respect to all
three cylinders $\Gamma_\alpha$,
$
\Gamma_{\alpha}=\{X\!\in\!{{\Bbb R}}^{6}:\,
X=({\bf x}_{\alpha}, {\bf y}_{\alpha}),\,
|{\bf x}_{\alpha}|=c_{\alpha}\}
$,
$\alpha=1,2,3$, where $c_{\alpha}>0$, stands for the values of  
$|{\bf x}_{\alpha}|$ when the cores of the particles in the pair 
$\alpha$ contact each other. A three-dimensional image of this 
space for particles with equal masses and the same core radii 
$c_\alpha=c$, $\alpha=1,2,3$, $c>0$, is sketched in Fig.~1, in 
coordinates $x_\alpha=|{\bf x}_\alpha|$, $y_\alpha=|{\bf 
y}_\alpha|$, and
$\eta_\alpha=\hat{\bf x}_\alpha{\cdot}\hat{\bf y}_\alpha$.  The
cylinders $\Gamma_1$, $\Gamma_2$ and $\Gamma_3$ are depicted in
this figure by the plane $x_1=c$ and surfaces $\frac{1}{4}
x_1^2+\frac{3}{4} y_1^2-$$\frac{\sqrt{3}}{2} x_1 y_1
\eta_1=c^2,\ $ $\frac{1}{4} x_1^2+\frac{3}{4}
y_1^2+$$\frac{\sqrt{3}}{2} x_1 y_1 \eta_1=c^2$, respectively.
The domain $\Omega$ is a part of the set $x_1>c$, $y_1>0$,
$-1\leq\eta_1\leq 1,$ restricted by $\Gamma_2$ and $\Gamma_3$.
{} From a geometrical point of view, the image shown in Fig.~1 is
exact%
%%%%%%%%%%%%%%%%%%%%%%%%%%%%%%%%%%%%%%%%%%%%%%%%%%%%%%%%%%%%%%%%
\footnote{It should be noted that the transition to the
variables $x_\alpha$, $y_\alpha$ and $\eta_\alpha$ is not
conformal. In particular the true angle between any two surfaces
$\Gamma_\beta$ and $\Gamma_\gamma$, $\beta\neq\gamma$, at points
belonging to the intersection manifold
$\Gamma_\beta\bigcap\Gamma_\gamma$
varies between
$\frac{\pi}{2}-\phi_{\beta\gamma}$ and
$\frac{\pi}{2}+\phi_{\beta\gamma}$,
$\phi_{\beta\gamma}=\arcsin|{\sf c}_{\beta\gamma}|$ and never
acquires the value of $0$ or $\pi$ (since $|{\sf
c}_{\beta\gamma}|<1$).}
%%%%%%%%%%%%%%%%%%%%%%%%%%%%%%%%%%%%%%%%%%%%%%%%%%%%%%%%%%%%%%%%
since only coordinates (such as Eulerian angles) describing a
rotation of the plane defined by the position of particles are
omitted.

The Hamiltonian of a system of three particles with hard-core
interactions is defined in $L_2(\Omega)$ by the expression
\begin{equation}
\label{Schr}
		Hf(X)=\left(-\Delta_X+\sum_{\alpha=1}^3
		V_{\alpha}\right)f(X)
\end{equation}
on the set of functions $f(X)$, $f\in W_2^2(\Omega)$, satisfying
the condition
\begin{equation}
\label{bccor}
	 	f\left.\right|_{\partial\Omega}=0
\end{equation}
on the boundary $\partial\Omega$ of the domain $\Omega$.  The
Laplacian $-\Delta_X$  corresponds to the kinetic energy operator
of the system under consideration. The potentials $V_\alpha$,
$\alpha=1,2,3$ are two-body interactions and thus when acting on
the function $f$ in the expression (\ref{Schr}) they only
operate on the corresponding two-body variable ${\bf
x}_{\alpha}$, $|{\bf x}_\alpha|>c_\alpha$.  We assume that these
pair potentials are bounded Hermitian operators. The Hamiltonian
$H$ that includes such potentials is a self-adjoint operator and
thus its  spectrum is real. For local potentials we assume that
\begin{equation}
\label{Vcond}
	        |V_\alpha({\bf x}_\alpha)|
       		\leq C_\alpha(1+|{\bf x}_\alpha|)^{-3-\varepsilon}\,,
          	\qquad |{\bf x}_\alpha|\geq c_\alpha\,,
\end{equation}
where  the constants $C_\alpha>0$  and $\varepsilon>0$. Similar
conditions are assumed for the partial derivatives of
$V_\alpha({\bf x}_\alpha)$. The Aziz {\it et al.} potentials
\cite{Aziz79,Aziz87} considered in this work are  examples of
such interactions. Nonlocal potentials can also be included in
our formalism, provided their kernels $V_\alpha({\bf
x}_\alpha,{\bf x'}_\alpha)$ are smooth functions  obeying
conditions similar to (\ref{Vcond}) as $|{\bf x}_{\alpha}|,|{\bf
x'}_{\alpha}|\rightarrow\infty$.

\subsection{Bound state problem}
%%%%%%%%%%%%%%%%%%%%%%%%%%%%%%%%%%
We shall consider first the boundary value problem for
the Faddeev differential equations for the  three-body bound
state. Let
$$
	      H\Psi=E\Psi\,,
$$
$E$ being the bound state energy and $\Psi$  the corresponding
three-body bound state wave function.  We are concerned with
states for which $E<0$ and that these energies are below the
threshold of the  continuous spectrum of $H$. Using the Green's
formula (see, e.g., Ref.  \cite{Potential}) one can  easily show
that the function $\Psi$ satisfies the following
Lippmann-Schwinger type equation
\begin{equation}
\label{LSchCor}
		\Psi(X)=-\displaystyle\int\limits_{\partial\Omega}
	       	d\sigma_S\, G_0(X,S;E)\frac{\partial}{\partial n_S}\Psi(S)
	      	-\displaystyle\sum_{\alpha=1}^3\,
	       	\displaystyle\int\limits_{\Omega} dX'\, G_0(X,X';E)
	       	(V_{\alpha}\Psi)(X')
\end{equation}
where  $G_0(X,X';z)$, is the three-body free Green function,
i.e., the kernel of the resolvent
$$
		G_0(z)=(-\Delta_X-z)^{-1}
$$
of the Laplacian $-\Delta_X$ in the six-dimensional space
${{\Bbb R}}^6$. We recall that the function $G_0(X,X'; z)$ can be
expressed in terms of the Hankel function of the first kind
$H^{(1)}_2$
$$
	        G_0(X,X';z)=\frac{{\rm i}z}{16\pi^2}\,
	  	\frac{H_{2}^{(1)} (\sqrt{z}|X-X'|)}{|X-X'|^2}\,.
$$
The $n_S$ denotes the external  unit vector (directed into
$\Omega$) normal to the surface $\partial\Omega$  while
$d\sigma_S$ is a surface element (five-dimensional square) on
$\partial\Omega$.

In contrast to $\Psi(X)$, defined only for $X\in\Omega$, the
function $G_0(X,X';E)$ is defined for all $X\in{{\Bbb R}}^6$, $X\neq
X'$. Therefore, the right-hand side of (\ref{LSchCor}) is
defined for  $X\in\Omega$ as well as for
$X\in{{\Bbb R}}^6\setminus\overline{\Omega}$. Moreover, from the
Green's formula  it follows that, for any
$X\in{{\Bbb R}}^6\setminus\overline{\Omega}$
\begin{equation}
\label{InsideOmega}
	  	-\displaystyle\int\limits_{\partial\Omega}d\sigma_S\, G_0(X,S;E)
	  	\frac{\partial}{\partial n_S} \Psi(S)
	  	-\displaystyle\sum_{\alpha=1}^3\,
          	\displaystyle\int\limits_{\Omega} dX'\, G_0(X,X';E)\,
	  	(V_\alpha\Psi)(X')=0 \,.
\end{equation}
The Faddeev components of the function $\Psi$ are introduced via
the formulas (see  Refs.  \cite{MMYa,MotovilovVLGU,EChAYa})
\begin{equation}
\label{FaddComp}
	  	\Phi_{\alpha}(X)=-\displaystyle\int\limits_{\Gamma_{\alpha}
                \bigcap\partial\Omega}d\sigma_S\, G_0(X,S;E)
	  	\frac{\partial}{\partial n_S} \Psi(S)
          	- \displaystyle\int\limits_{\Omega} dX'\, G_0(X,X';E)\,
	 	( V_\alpha\Psi)(X')\,.
\end{equation}
We shall consider the functions $\Phi_{\alpha}(X)$ given by
(\ref{FaddComp}) for all $X\in{{\Bbb R}}^6$, i.~e.,  outside as well
as inside the surface $\partial\Omega$. From (\ref{LSchCor}) and
(\ref{InsideOmega}) one gets
\begin{equation}
\label{SummFadd}
       \sum_{\alpha=1}^{3} \Phi_{\alpha}(X)=\left\{\begin{array}{cl}
    \Psi(X), & X\in\Omega \\0, & X\in{\Bbb R}^6\setminus\overline{\Omega}\,.
\end{array}\right.
\end{equation}
The surface integral
\begin{equation}
\label{PotSimpleLayer}
	   	\displaystyle\int\limits_{\Gamma_{\alpha}} d\sigma_S\,
                G_0(X,S;z)\,\mu_{\alpha}(S)\,,\quad z\in{\Bbb C}\,,
\end{equation}
which appears in (\ref{FaddComp}), represents the potential of a
simple layer \cite{Potential} with density $\mu_{\alpha}$
concentrated on the surface $\Gamma_{\alpha}$. In our case
$$
    \mu_{\alpha}(S)=
\cases{
    &$\displaystyle\frac{\partial}{\partial n_S}\Psi(S)$ \qquad if
		  $S\in\Gamma_{\alpha}\bigcap\partial\Omega $\cr\cr
     & \quad $0$ \qquad \qquad \quad
                 if $S\in\Gamma_{\alpha}\setminus\partial\Omega$\,.\cr}
$$
As has been shown in Refs.~\cite{MerMot,MotovilovVLGU,MM-Kalinin},
each of the densities
$\mu_{\alpha}(S)$  on the cylinder $\Gamma_{\alpha}$ as a
function of the variable $S\in \Gamma_\alpha$, is everywhere
continuous except perhaps where this  cylinder intersects the
other two cylinders $\Gamma_{\beta}$, $\beta\neq\alpha$. This
means that in crossing the surface $\Gamma_{\alpha}$ (at least
not on the intersection of $\Gamma_\alpha$ with $\Gamma_\beta$),
the potential of a simple layer (\ref{PotSimpleLayer}) is a
continuous function \cite{Potential}.  Evidently for
$X\not\in\Gamma_{\alpha}$ the integral (\ref{PotSimpleLayer}) is
infinitely differentiable with respect to $X$ and that
$$
      		(-\Delta_X-z)\displaystyle\int\limits_{\Gamma_{\alpha}}
	  	d\sigma_S\, G_0(X,S;z)\,\mu_{\alpha}(S)=0\,.
$$
Acting on both sides of the equality (\ref{FaddComp}) by the
differential expression $-\Delta_X-E$ and taking into account
the relation (\ref{SummFadd}), one obtains the following system
of differential equations for the components $\Phi_{\alpha}(X)$,
\begin{equation}
\label{FaddeevEq}
\left\{
    \begin{array}{rcll}
	 	(-\Delta_X+V_{\alpha}-E)\Phi_{\alpha}(X)
		&=&-V_{\alpha} \displaystyle\sum
		\limits_{\beta\neq\alpha}\Phi_{\beta}(X)\,,
		\, \, & \,\, |{\bf x}_{\alpha}|>c_{\alpha}\,, \\
	 	(-\Delta_X-E)\Phi_{\alpha}(X)
		 &=& 0\,, \,\, & \,\, |{\bf x}_{\alpha}|<c_{\alpha}\,.
     \end{array}\right.
\end{equation}
According to (\ref{SummFadd}), the sum of the functions
$\Phi_{\alpha}(X)$ must vanish not only on the surface
$\partial\Omega$ but also inside of it, i.e.,
\begin{equation}
\label{SummFaddInside}
	 	\displaystyle\sum_{\beta=1}^3 \Phi_{\beta}(X)\equiv
                0, \quad X\in{{\Bbb R}}^6\setminus{\Omega}\,.
\end{equation}
In fact one can replace the very strong conditions
(\ref{SummFaddInside}) with the essentially more weak
conditions \cite{MerMot,MotovilovVLGU}
\begin{equation}
\label{SummFaddCylinder}
       \left. \displaystyle\sum_{\beta=1}^3 \Phi_\beta(X)\right |_{
          |{\bf x}_\alpha|=c_\alpha}=0, \qquad \alpha=1,2,3\,,
\end{equation}
requiring that the sum of $\Phi_{\alpha}(X)$ to be zero only on
the cylinders $\Gamma_{\alpha}$.  It is understood that for the
bound-state problem, the conditions
\begin{equation}
\label{L2}
                \Phi_{\alpha}\in L_2({{\Bbb R}}^6), \qquad \alpha=1,2,3\,,
\end{equation}
must be fulfilled. Similarly to the pure potential
model~\cite{MF} the asymptotic behaviour of $\Phi_\alpha(X)$ as
$|X|\to\infty$ is of an exponential character, the form of which
is quite complicated~\cite{MF}.

Equations  (\ref{FaddeevEq}), (\ref{SummFaddCylinder}), and
(\ref{L2}) describe the boundary value problem  for three-body
bound systems with hard-core interactions and are a natural
generalization of the Faddeev differential  formulation
\cite{MF} for bound states.

The numerical advantage of our approach is already obvious from
the structure of Eqs.  (\ref{FaddeevEq}):  When a potential with
a strong repulsive core is replaced with the  hard-core model,
one approximates, inside the core domains, only the Laplacian
$-\Delta_X$, instead of the sum of the Laplacian and a huge
repulsive term, and in this way a much better numerical
approximation can be achieved.

\subsection{Scattering processes}
%%%%%%%%%%%%%%%%%%%%%%%%%%%%%%%%%%
%
Let $\Psi^{[\beta,\xi]\pm}(X,{\bf p}_\beta )$  be the three-body 
wave function corresponding to a ($2+1\rightarrow 
2+1\,;\,1+1+1$) process where in the  initial state the  pair 
subsystem $\beta$ is bound in a state $\psi_{\beta,\xi}({\bf 
x}_\beta)$ with energy $\epsilon_{\beta,\xi}$, 
$\epsilon_{\beta,\xi}<0$, and the complementary particle is 
asymptotically free, the relative momentum  being ${\bf 
p}_\beta$, ${\bf p}_\beta\in{{\Bbb R}}^3$.   By $\xi$ we denote 
here a distinctive label (consisting of appropriate quantum 
numbers) for the two-body state concerned.  The Faddeev 
components \cite{MotovilovVLGU,EChAYa} 
$\Phi_\alpha(X)\equiv\Phi_\alpha^{[\beta,\xi]\pm}(X,{\bf 
p}_\beta)$ of the wave function $\Psi^{[\beta,\xi]\pm}(X,{\bf 
p}_\beta)$,
$$
    		\Psi^{[\beta,\xi]\pm}(X)=\sum_{\alpha=1}^3
    		\Phi^{[\beta,\xi]\pm}_\alpha(X)\,,
$$
in the hard-core model satisfy the same differential
equations~(\ref{FaddeevEq}) and boundary
conditions~(\ref{SummFaddCylinder}) of the three-body bound
state problem.  These components can be written as
\begin{equation}
\label{Repr}
             \Phi_\alpha^{[\beta,\xi]\pm}(X,{\bf p}_\beta)=
             \delta_{\alpha\beta} \chi_{\beta,\xi}(X,{\bf p}_\beta)
             +\displaystyle\sum_{\xi'}\psi_{\alpha,\xi'}({\bf x}_\alpha)
             U_{\alpha,\xi'}^{[\beta,\xi]\pm}({\bf y}_\alpha,{\bf p}_\beta)
             +U_{\alpha,0}^{[\beta,\xi]\pm}(X,{\bf p}_\beta)
\end{equation}
where
\begin{equation}
\label{Incident}
    \chi_{\beta,\xi}(X,{\bf p}_\beta)=\psi_{\beta,\xi}({\bf x}_\beta)
       \exp({\rm i}\,{\bf p}_\beta{{\cdot}}{\bf y}_\beta)
\end{equation}
is the incident wave consisting of a two-body bound state
$\psi_{\beta,\xi}$ and a plane wave.  The functions
$U_{\alpha,\xi'}^{[\beta,\xi]\pm}$ and
$U_{\alpha,0}^{[\beta,\xi]\pm}$ have the same asymptotic
behavior~\cite{MotovilovThesis} as in the usual potential model
\cite{MF}, namely,
\begin{eqnarray}
\label{As2}
 U_{\alpha,\xi'}^{[\beta,\xi]\pm}({\bf y}_\alpha,{\bf p}_\beta)
   & \begin{array}{c}
    \phantom{X^a}\\
    {\mbox{\Large$=$} }\\
    \mbox{\scriptsize$y_\alpha\rightarrow\infty$}
    \end{array}&
  \displaystyle\frac{{\rm e}^{\pm {\rm i}\sqrt{E-\epsilon_{\alpha,\xi'}}
            |{\bf y}_\alpha|}} {|{\bf y}_\alpha|}
             \left[{\rm a}_{\alpha,\xi'}^{[\beta,\xi]\pm}
            (\hat{\bf y}_\alpha,{\bf p}_\beta)
  +o(|{\bf y}_\alpha|^{-1/2})\right]\,, \\
\label{As3}
       U_{\alpha,0}^{[\beta,\xi]\pm}(X,p_\beta)
       & \begin{array}{c}
    \phantom{X^a}\\
    {\mbox{\Large$=$} }\\
    \mbox{\scriptsize$X\rightarrow\infty$}
    \end{array} &
        \displaystyle\frac{{\rm e}^{\pm{\rm i}\sqrt{E}|X|}}{|X|^{5/2}}
  \left[A_{\alpha}^{[\beta,\xi]\pm}(\hat{X},
   {\bf p}_\beta)+o(|X|^{-1/2})\right]
\end{eqnarray}
where $E=\epsilon_{\beta,\xi}+p_\beta^2$, is the  energy of the
system. For $E>\epsilon_{\alpha,\xi'}$ the function
$
{\rm a}^{[\beta,\xi]\pm}_{\alpha,\xi'}(\hat{\bf y}_\alpha,{\bf p}_\beta)
$
represents the amplitude for the elastic ($\alpha=\beta$,
$\xi'=\xi$) or rearrangement ($\alpha\neq\beta$ or $\xi'\neq
\xi$) scattering.  The functions $A^{[\beta,\xi]\pm}_\alpha
(\hat{X},{\bf p}_\beta)$ provide us, at $E>0$, with the Faddeev
components of the total breakup amplitude $ {\cal
A}^{[\beta,\xi]\pm}(\hat{X},{\bf p}_\beta)$
$$
      		{\cal A}^{[\beta,\xi]\pm}(\hat{X},{\bf p}_\beta)=
       		\displaystyle\sum_{\alpha=1}^{3}
       		A^{[\beta,\xi]\pm}_\alpha(\hat{X},{\bf p}_\beta)\,.
$$
It should be stressed that the two-body eigenfunctions
$\psi_{\alpha,\xi}({\bf x}_\alpha)$ are assumed to be zero
within the respective cores $\alpha$, i.e.,
$\psi_{\alpha,\xi}({\bf x}_\alpha)\equiv 0$ for $|{\bf
x}_\alpha|\leq c_\alpha$.  The boundary-value problem  as
described by Eqs. (\ref{FaddeevEq}), (\ref{SummFaddCylinder}),
and (\ref{Repr})--(\ref{As3}), is the extension of the Faddeev
formalism to the $(2+1\rightarrow 2+1\,;\,1+1+1)$ scattering
processes for hard-core potentials.

A detail analysis for the boundary-value problems described
above, the derivation of the asymptotic boundary conditions for
scattering states as well as other boundary-value formulations,
can be found in Refs. \cite{MotovilovThesis,EChAYa}. Here, we
only recall, briefly, some peculiar properties of the discrete
spectrum generated by the condition~(\ref{SummFaddCylinder}).
As compared to the spectrum of the initial Hamiltonian defined
by Eqs.~(\ref{Schr}) and~(\ref{bccor}), this spectrum acquires
an additional component, corresponding to the Dirichlet
boundary-value problems for the domains which result from the
intersection of the cylinders $\Gamma_\alpha$.  We introduce the
following notations for these domains:  Let
$\Lambda_{\alpha\beta\gamma}$ be a domain restricted by all the
three cylinders $\Gamma_\alpha$, $\alpha=1,2,3,$ and
$\partial\Lambda_{\alpha\beta\gamma}$ be its boundary  (see
Fig.~1).  The notation $\Lambda_{\alpha\beta}$ is used
for a part of the domain bounded by the cylinders
$\Gamma_\alpha$ and $\Gamma_\beta$, $\beta\neq\alpha,$ and at
the same time is external with respect to the set
$\Lambda_{\alpha\beta\gamma}$.  By $\Lambda_\alpha$ we denote
the  domain bounded by the cylinder $\Gamma_\alpha$ which is at
the same time external to the rest cylinders $\Gamma_\beta$,
$\beta\neq\alpha$. The notations $\partial\Lambda_{\alpha\beta}$
and $\partial\Lambda_\alpha$ are used for the boundaries of the
domains $\Lambda_{\alpha\beta}$ and $\Lambda_\alpha$,
respectively.

It can be shown~\cite{MotovilovThesis} that the discrete
spectrum of the boundary-value problem (\ref{FaddeevEq}),
(\ref{SummFaddCylinder}), and (\ref{L2}) includes not only the
discrete spectrum $\sigma_d(H)$ of the original Hamiltonian $H$
but also a set $\sigma_d^{\rm aux}$ consisting of a discrete set
of eigenvalues of the homogeneous internal Dirichlet problems in
the domains $\Lambda_{\alpha\beta\gamma}$,
$\Lambda_{\alpha\beta}$, and $\Lambda_\alpha$, $\alpha,
\beta=1,2,3,$ $\beta\neq\alpha$, namely, the discrete spectra of
the operators defined in $W_2^2(\Lambda_{\alpha\beta\gamma})$,
$W_2^2(\Lambda_\alpha)$, and $W_2^2(\Lambda_{\alpha\beta})$ by
 the expression (\ref{Schr}) (under the assumption $(V_\alpha
f)({\bf x}_\alpha)\equiv 0$ for $|{\bf x}_\alpha|<c_\alpha $)
and the respective boundary conditions
           $f\left.\right|_{\partial\Lambda_{\alpha\beta\gamma}}=0$,
           $f\left.\right|_{\partial\Lambda_{\alpha}}=0$ and
           $f\left.\right|_{\partial\Lambda_{\alpha\beta}}=0$.
There exists a simple criterion in selecting solutions of the
spectral problem described by (\ref{FaddeevEq}),
(\ref{SummFaddCylinder}), and (\ref{L2})  corresponding  to the
spectrum of the Hamiltonian $H$ only.  This is just the
requirement (see also the condition (\ref{SummFaddInside})) that
the total wave function inside the cylinders $\Gamma_\alpha$
vanishes,
$$
   \Psi(X)=\displaystyle\sum_{\beta=1}^3 \Phi_\beta(X)\equiv 0 \quad
      \mbox{if } |{\bf x}_\alpha|<c_\alpha, \quad\alpha=1,2,3\,.
$$
It should be noted that the lower boundary ${\rm
inf}\,\sigma_d^{\rm aux}$ of the auxiliary spectrum $\sigma_d^{\rm
aux}$ is situated above%
%%%%%%%%%%%%%%%%%%%%%%%%%%%%%%%%%%%%%%%%%%%%%%%%%%%%%%%%%%%%%%%
\footnote{In the case of sufficiently small
$c_\alpha$, ${\rm inf}\,\sigma_d^{\rm aux}$ is positive and
behaves as $1/c^2$ where $c=\begin{array}{c}
    \phantom{a}\\[-7pt]
    {\rm max}\,\\[-7pt] \mbox{\scriptsize$\alpha$}
    \end{array} c_\alpha$.}
%%%%%%%%%%%%%%%%%%%%%%%%%%%%%%%%%%%%%%%%%%%%%%%%%%%%%%%%%%%%%%%
the lower boundary of the spectrum of the Hamiltonian $H$.
Therefore in searching for a ground state no validity check of
this criterion is necessary.

The elements of the set $\sigma_d^{\rm aux}$ are  points where
the $(2+1\rightarrow 2+1 \, ;\,1+1+1)$ scattering problems
(\ref{FaddeevEq}), (\ref{SummFaddCylinder}),
(\ref{Repr})--(\ref{As3}) have no a unique solution
\cite{MotovilovThesis}.  However the auxiliary spectrum
$\sigma_d^{\rm aux}$ is discrete and thus in practice a
coincidence of the scattering energy $E$ with a point of the set
$\sigma_d^{\rm aux}$ can be considered as an exceptional case.
In principle there is a way to avoid such a coincidence namely
by shifting the spectrum $\sigma_d^{\rm aux}$. This can be made,
for example, by replacing the zero values of the potentials
$V_\alpha$ inside the core domains by appropriate positive
values. Such a replacement does not affect the total wave
function $\Psi(X)=\displaystyle\sum\limits_\alpha\Phi_\alpha(X)$
in the physical domain, that is, at $X\in\Omega.$

%%%%%%%%%%%%%%%%%%%%%%%%%%%%%%%%%%%%%%%%%%%%%%%%%%%%%%%%%%%%%%%%%%%%%
\section{Partial boundary-value problems}
\label{Bosons}
%%%%%%%%%%%%%%%%%%%%%%%%%%%%%%%%%%%%%%%%%%%%%%%%%%%%%%%%%%%%%%%%%%%%%
In what follows  we shall  concentrate on a system of three
identical bosons interacting via a central potential $V$, i.e.,
via $V_\alpha({\bf x}_\alpha)=V(|{\bf x}_\alpha|)$,
$\alpha=1,2,3$.  The total wave function of the system  is
invariant under the permutation of particles belonging to any
pair $\alpha$, $P_\alpha\Psi=\Psi$, where $P_\alpha$ is the
permutation operator.  This means that  $\Psi(-{\bf
x}_\alpha,{\bf y}_\alpha)= \Psi({\bf x}_\alpha,{\bf y}_\alpha)$,
\, $\alpha=1,2,3.$ Thus from the definition of the Faddeev
components (\ref{FaddComp}) one obtains
\begin{equation}
\label{PF}
		P_\alpha\Phi_\alpha=\Phi_\alpha
\end{equation}
i.~e.
\begin{equation}
		\Phi_\alpha(-{\bf x}_\alpha,{\bf y}_\alpha)=
          	\Phi_\alpha({\bf x}_\alpha,{\bf y}_\alpha) \,.
\end{equation}
Furthermore
\begin{equation}
\label{Cyclic}
       \Phi_\beta=P^+\Phi_\alpha, \quad \Phi_\gamma=P^-\Phi_\alpha
\end{equation}
where $P^\pm$ stand for operators of cyclic permutation  of particles
\begin{equation}
\label{Perm}
        	P^+(123)=(312), \quad P^-(123)=(231).
\end{equation}
The conditions (\ref{Cyclic}) mean that the total wave function
$\Psi(X)$  is written as
\begin{equation}
\label{WaveFun}
      		\Psi=(I+{P}^{+}+{P}^{-})\Phi_\alpha
\end{equation}
where $I$ is the identity operator. Similarly, the Faddeev
equations~(\ref{FaddeevEq})  and  the hard-core boundary
conditions~(\ref{SummFaddCylinder}) are written as
\begin{eqnarray}
\label{FNeq}
    (-\Delta_X+V_\alpha-E)\Phi_\alpha(X) &=&  -V_\alpha
        ({P}^{+}+{P}^{-})\Phi_\alpha(X)\,,\qquad
         	|{\bf x}_\alpha|>c_\alpha\,,\\
     \label{Gelm}(-\Delta_X-E)\Phi_\alpha(X) &=&  0\,,\qquad
   \qquad\qquad\qquad\qquad\quad |{\bf x}_\alpha|<c_\alpha\,,
\end{eqnarray}
and
\begin{equation}
\label{BosBC}
       \Phi_\alpha(X)=-({P}^{+}+{P}^{-}) \Phi_\alpha(X)\,,\qquad
                |{\bf x}_\alpha|=c_\alpha\,,
\end{equation}
where $c_1=c_2=c_3=c$ and, say, $\alpha=1$.  In what follows we
shall drop, for convenience, the identification  $\alpha$. If
one searches for a bound state of the system, the condition
\begin{equation}
\label{L2Bos}
                \Phi\in L_2({{\Bbb R}}^6)
\end{equation}
is required.

Consider now a $(2+1\longrightarrow 2+1\,;\,1+1+1)$ scattering
process for the three bosons in an initial state
\begin{equation}
\label{initialBos}
\chi_\xi(X,{\bf p})=\psi_\xi({\bf x})
\exp({\rm i}\,{\bf p}{\cdot}{\bf y})\,.
\end{equation}
Since the particles are identical, the incident wave
$\chi_\xi(X,{\bf p})$ must be included now, in contrast to
(\ref{Repr}), into all three summands of the total scattering
wave function $\Psi(X)\equiv\Psi^{\xi\pm}(X,{\bf p})$ given by
equation (\ref{WaveFun}) with $\Phi(X) \equiv\Phi^{\xi\pm}$.
Therefore the Faddeev components $\Phi^{\xi\pm}$  have the form
\begin{equation}
\label{ReprBos}
		\Phi^{\xi\pm}(X,{\bf p})=
      		\chi_\xi(X,{\bf p})+\displaystyle\sum_{\xi'}
       		\psi_{\xi'}({\bf x})U_{\xi'}^{\xi\pm}({\bf y},{\bf p})
       		+U_0^{\xi\pm}(X,{\bf p})
\end{equation}
where the terms $U_{\xi'}^{\xi\pm}$ and $U_0^{\xi\pm}$ have the
same asymptotic form as (\ref{As2}) and (\ref{As3}),
\begin{eqnarray}
\label{As2Bos}
   		U_{\xi'}^{\xi\pm}({\bf y,p}) &
          \begin{array}{c}\phantom{X^a}\\
       {\mbox{\Large$=$}\,}\\
       \mbox{\scriptsize${\bf y}\rightarrow\infty$}
           \end{array}&
           \displaystyle\frac{{\rm e}^{\pm{\rm i}\sqrt{E-\epsilon_{\xi'} }
              |{\bf y}|}}{|{\bf y}|}
   		\left[{\rm a}_{\xi'}^{\xi\pm} (\hat{\bf y},{\bf p})+
   		o(|{\bf y}|^{-1/2})\right]\,,\\
\label{As3Bos}
   		U_{0}^{\xi\pm}(X,{\bf p}) &
              \begin{array}{c}\phantom{X^a}\\
       {\mbox{\Large$=$}\,}\\
       \mbox{\scriptsize$X\rightarrow\infty$}
           \end{array} &
         \displaystyle\frac{{\rm e}^{\pm{\rm i}\sqrt{E}|X|}}{|X|^{5/2}}
   		\left[A^{\xi\pm}(\hat{X},{\bf p}) +
   		o(|X|^{-1/2})\right]
\end{eqnarray}
where $E=\epsilon_\xi+|{\bf p}|^2$. If $E>\epsilon_{\xi'}$, the
function ${\rm a}_{\xi'}^{\xi\pm} (\hat{\bf y},{\bf p})$
represents  the elastic scattering  amplitude, $\xi'=\xi$, or
the rearrangement one, $\xi'\neq \xi$. At $E>0$ the function
$A^\xi$ represents the Faddeev component of the total breakup
amplitude ${\cal A}^{\xi\pm}(\hat{X},{\bf p})$ which is
expressed via $A^{\xi\pm}$
$$
     		{\cal A}^{\xi\pm}(\hat{X},{\bf p})=
		\left(I+P^+ +P^-\right)A^{\xi\pm}(\hat{X},{\bf p})\,.
$$
The description for the auxiliary spectrum $\sigma_d^{\rm aux}$
of the boundary-value problems (\ref{FNeq})--(\ref{BosBC}),
(\ref{L2Bos}) and~(\ref{FNeq})--(\ref{BosBC}),
(\ref{ReprBos})--(\ref{As3Bos}) is the same as the one outlined
in Sect.~II except that all core sizes are now equal,
$c_\alpha=c$, $\alpha=1,2,3$.  Further we consider the case of
the $\Psi^{\xi+}$ scattering wave functions and thus the index
``$+$'' will be omitted.

Similarly to Eqs.~(\ref{FaddeevEq}) and
(\ref{SummFaddCylinder}), Eqs.  (\ref{FNeq})--(\ref{BosBC})  are
six-dimensional. Therefore we may use,  for their partial wave
expansion, the bispherical basis
\begin{equation}
\label{basis}
           	|l\lambda L\rangle=\displaystyle\sum\limits_{m+\mu=M}
           	\langle lm\lambda\mu|LM\rangle \,\, Y_l^m(\hat{\bf x})
            	Y_\lambda^\mu(\hat{\bf y})
\end{equation}
where $L$ is the total angular momentum of the system,
$Y_l^m(\hat{\bf x})$ and $Y_\lambda^\mu(\hat{\bf y})$, are the
spherical harmonics, and $\langle lm\lambda\mu|LM\rangle$ the
Clebsch--Gordan coefficients.

The potential $V$, being central, is diagonal in the basis
(\ref{basis}) and has the same diagonal elements in all partial
waves.  Since the operator of the total angular momentum $\bf L$
and its projection ${\bf L}_z$ commute with both the Laplacian
$-\Delta_X$ and the sum $I+{P}^{+}+{P}^{-}$, the study of the
boundary-value problems (\ref{FNeq})--(\ref{BosBC}),
(\ref{L2Bos}),  and (\ref{FNeq})--(\ref{BosBC}),
(\ref{ReprBos})--(\ref{As3Bos}) is reduced to a study in
subspaces corresponding to fixed values of the momentum $L$ and
its projection $M$.  Since the index $M$  does not effect the
structure of the equations it will be omitted. Thus  $\Phi_L(X)$
denotes the  partial components of  $\Phi(X)$.

Expanding the function $\Phi_L(X)$ in a series of bispherical harmonics
\begin{equation}
\label{bisph}
       		\Phi_L(X)=\displaystyle\sum\limits_a
       		\displaystyle\frac{\Phi_{aL}(x,y)}{x\,y}\,|aL\rangle\,,
        	\quad a=\{l,\lambda\}\,,
         	\quad x=|{\bf x}|\,\mbox{ and }\, y=|{\bf y}|\,,
\end{equation}
and using the  results of Ref.~\cite{MGL}  (see also
\cite{MF,EChAYa}) one obtains  for (\ref{FNeq})--(\ref{Gelm})
the following partial equations
\begin{equation}
\label{FadPart}
            	(H_L-E)\Phi_{aL}(x,y)=\left\{
            	\begin{array}{cl} -V(x)\Psi_{aL}(x,y), & x>c \\
                        0,                  & x<c
\end{array}\right.
\end{equation}
where
$$
		H_L=-\displaystyle\frac{\partial^2}{\partial x^2}
    		-\displaystyle\frac{\partial^2}{\partial y^2}
    		+\displaystyle\frac{l(l+1)}{x^2}
    		+\displaystyle\frac{\lambda(\lambda+1)}{y^2}\,.
$$
The function $\Psi_{aL}(x,y)$ represents the partial component
of the total wave function (\ref{WaveFun}) and is related to the
partial Faddeev components $\Phi_{aL}(x,y)$ by
\begin{equation}
\label{FTconn}
            	\Psi_{aL}(x,y)=\Phi_{aL}(x,y) + \sum_{a'}\int_{-1}^{+1}
                d\eta\,h_{a a'}^L(x,y,\eta)\,\Phi_{a'L}(x',y')
\end{equation}
where
$$
        	x'=\sqrt{\displaystyle\frac{1}{4}\,x^2+\displaystyle
          \frac{3}{4}\,y^2-\displaystyle \frac{\sqrt{3}}{2}\,xy\eta}\,,
$$
$$
          	y'=\sqrt{\displaystyle\frac{3}{4}\,x^2+\displaystyle
         \frac{1}{4}\,y^2+ \displaystyle\frac{\sqrt{3}}{2}\,xy\eta}\,,
$$
with $\eta=\hat{\bf x}{\cdot}\hat{\bf y}$. The functions $h_{aa'}^L$
are given by \cite{MGL} (see also \cite{MF})
\begin{eqnarray}
          	h_{aa'}^L & =& \displaystyle\frac{xy}{x'y'} \,
          	(-1)^{l+L} \,\frac{(2\lambda+1)(2l+1)}{2^{\lambda+l}}
          	\left[(2\lambda)! (2l)! (2\lambda'+1) (2l'+1)\right]^{1/2}
\nonumber\\
          &\times& \sum\limits_{k=0}^{k_{max}} (-1)^k (2k+1) P_k (\eta)
       \sum_{\stackrel{\lambda_1 + \lambda_2 = \lambda,}{l_1 + l_2 = l}}
           	\displaystyle\frac{y^{\lambda_1 + l_1}
           	x^{\lambda_2 + l_2}}{y'^{\lambda}
           	x'^{l}} (-1)^{l_2} (\sqrt{3})^{\lambda_2 +l_1}
 \nonumber \\
           	&\times& \left[(2\lambda_1)! (2l_1)! (2\lambda_2)!
         (2l_2)!\right]^{-1/2} \sum_{\lambda'' l''} (2\lambda''+1)(2l''+1)
       		\left(
\begin{array}{ccc}
                       \lambda_1 & l_1 & \lambda'' \\
                        0 & 0 & 0
\end{array}
       \right) \\
&\times&
        \left(
\begin{array}{ccc}
 			\lambda_2 & l_2 & l'' \\
			0 & 0 & 0
\end{array}\right)
	\left(
\begin{array}{ccc}
			 k & \lambda'' & \lambda' \\
			 0 & 0 & 0
\end{array}\right)
	\left(
\begin{array}{ccc}
			k & l'' & l' \\
			0 & 0 & 0
\end{array}
        \right)
	\left\{
\begin{array}{ccc}
			 l' & \lambda' & L\\
			\lambda'' & l'' & k
\end{array}
	\right\}
	\left\{
\begin{array}{ccc}
			 \lambda_1 & \lambda_2 & \lambda\\
			 l_1 & l_2 & l \\
		         \lambda'' & l'' & L
\end{array}
  	\right\}\, ,
\nonumber \\
	k_{max} &=& \displaystyle\frac{l+\lambda+l'+\lambda'}{2},
 \nonumber
\end{eqnarray}
where $P_k(\eta)$ is the Legendre polynomial of order $k$. In the
above, the standard notation for the 3-$j$, 6-$j$, and 9-$j$
Wigner symbols, as defined in \cite{Messiah}, is used. It should
be noted that the kernels $h^L_{aa'}$  depend  only on  the
hyperangles
\begin{equation}
\label{thetas}
\theta=\arctan\frac{y}{x}\quad
\mbox{and}\quad \theta'=\arctan\frac{y'}{x'}
\end{equation}
and not on the hyperradius
\begin{equation}
\label{rhos}
   		\rho=\sqrt{x^2+y^2}=\sqrt{x'^2+y'^2}.
\end{equation}
Due to (\ref{PF}) only the components $\Phi_{aL}$ corresponding
to $a=\{l,\lambda\}$ with even $l$ are unequal to zero. This
reduces considerably the number of coupled equations to be
solved.

The functions $\Phi_{aL}(x,y)$ satisfy the boundary conditions
\begin{equation}
\label{BCStandard}
		\Phi_{aL}(x,y)\left.\right|_{x=0}=0\quad\mbox{\rm and}\,\,
		\Phi_{aL}(x,y)\left.\right|_{y=0}=0\,.
\end{equation}
The partial  wave version of the hard-core conditions
(\ref{BosBC}) is given by $\Psi_{aL}(x,y)
\left.\right|_{x=c}=0$, that is, by
\begin{equation}
\label{BCCorePart}
		\Phi_{aL}(c,y) + \sum_{a'}\int_{-1}^{+1}
           d\eta\,h_{a a'}^L(c,y,\eta)\,\Phi_{a'L}(x',y')=0\,.
\end{equation}

For the  bound-state problem one requires that the functions
$\Phi_{aL}(x,y)$ are square integrable in the quadrant $x\geq
0$, $y\geq 0$, i.e., they must satisfy the condition
$\Phi_{aL}\in L_2({{\Bbb R}}^2_+)$ which follows from (\ref{L2Bos}).
A more detailed and useful in bound state calculations is the
asymptotic condition
\begin{eqnarray}
\Phi_{aL} &=& \sum\limits_\nu \psi_{l,\nu}(x)\,
h_\lambda(\sqrt{E-\epsilon_{l,\nu}}\,y)
\left[{\rm a}_{aL,\nu}+o(y^{-1/2})\right]\nonumber\\
\label{BScond}
&+& \frac{\exp({\rm i}\sqrt{E}\rho+i\pi L/2)}{\sqrt{\rho}}
\left[A_{aL}(\theta)+o(\rho^{-1/2})\right]
\end{eqnarray}
where $E$ ($E<0$) is the bound-state energy, $a=\{l,\lambda\}$, 
and $\psi_{l,\nu}(x)$ is the two-body partial wave function 
corresponding to a $\nu$-th bound state $\epsilon_{l,\nu}$ for 
the angular momentum value $l$. Here $h_{\lambda}$ is used for 
the spherical Hankel function. The coefficients ${\rm 
a}_{aL,\nu}$ and $A_{aL}(\theta)$ describe contributions into 
$\Phi_{aL}$ (and $\Psi_{aL}$) from the  $(2+1)$ and $(1+1+1)$ 
channels respectively.  The formula~(\ref{BScond}) follows from 
the asymptotic expression of the total Faddeev component of 
the bound-state wave function (see Ref. ~\cite{MF}, Chapter~IV, final 
subsection of \S 3) which is also valid for the hard-core model.

The asymptotic boundary conditions for the partial Faddeev
components of the $(2+1\rightarrow 2+1\,;\,1+1+1)$ scattering
wave function as $X\rightarrow\infty$ and/or ${\bf y}\rightarrow
\infty$ follow from~(\ref{ReprBos})--(\ref{As3Bos}).  These are
\begin{equation}
\label{AsBCPart}
\begin{array}{rcl}
		\Phi_{a'L}^{[a,\nu]}(x,y,p) & = &
		\delta_{a'a}\psi_{l,\nu}(x)j_\lambda(py) \\
	 	&+& \displaystyle\sum\limits_{\nu'} \psi_{l',\nu'}(x)
		h_{\lambda'}(\sqrt{E-\epsilon_{l',\nu'}}\,y)
		\left[{\rm a}_{a'L,\nu'}^{[a,\nu]}(p)+o\left(y^{-1/2}
		\right)\right] \\
 &+& \displaystyle\frac{\exp({\rm i}\sqrt{E}\rho+
{\rm i}\pi L/2)}{\sqrt{\rho}}
\left[A_{a'L}^{[a,\nu]}(p,\theta)+o\left(\rho^{-1/2}\right)\right]
\end{array}
\end{equation}
where $p=|{\bf p}|$ is the relative moment conjugate to the
Jacobi variable $y$ and the scattering energy $E$ is given by
$E=\epsilon_{l,\nu}+p^2$.  The  $j_{\lambda'}$ stands for the
spherical Bessel function.  The value ${\rm
a}^{[a,\nu]}_{a'L,\nu'}$ represents, at $E> \epsilon_{l',\nu'}$,
the partial amplitude of an elastic scattering, $a'=a$ and
$\nu'=\nu$, or rearrangement,  $a'\neq a$ or $\nu'\neq\nu$,
process.  The functions $A_{a'L}^{[a,\nu]}(\theta)$ provide us,
at $E>0$, the corresponding partial Faddeev breakup amplitudes.
Finally the  physical partial breakup amplitudes  are written as
\begin{equation}
            {\cal A}^{[a,\nu]}_{a'L}(\theta) = A^{[a,\nu]}_{a'L}(\theta)+
            \sum\limits_{a''} \int_{-1}^{1}
         d\eta\, h^L_{a'a''}(x,y,\eta)\,A^{[a,\nu]}_{a''L}(\theta')
\end{equation}
where $\theta$ and $\theta'$ are given by (\ref{thetas}).

%%%%%%%%%%%%%%%%%%%%%%%%%%%%%%%%%%%%%%%%%%%%%%%%%%%%%%%%%%%%%
%%%%%%%%%%%%%%%%%% SECT IV   %%%%%%%%%%%%%%%%%%%%%%%%%%%%%%%%%
%%%%%%%%%%%%%%%%%%%%%%%%%%%%%%%%%%%%%%%%%%%%%%%%%%%%%%%%%%%%%%
\section{Application to the three-atomic
         $\mbox{$^{\bf 4}${\bf H\lowercase{e}}}$ system}
\label{results}
%%%%%%%%%%%%%%%%%%%%%%%%%%%%%%%%%%%%%%%%%%%%%%%%%%%%%%%%%%%%%%%%%%%%%
We employed the Faddeev equations (\ref{FadPart}) and the
hard-core boundary condition (\ref{BCCorePart}) to calculate the
binding energies of the Helium atomic trimer and the ultra--low
energy phase shifts of the Helium atom scattered off the Helium
diatomic molecule.  As a $^4$He--$^4$He interatomic interactions
we use the HFDHE2 \cite{Aziz79} and HFD-B \cite{Aziz87}
potentials of Aziz and co-workers.
Both HFDHE2 and HFD-B potentials have the form%
\begin{equation}
\label{Aziz1-2}
	V(r)=\varepsilon \left \{ A\exp(-\alpha\zeta
	+\beta\zeta^2 ) -\left [ \frac{C_6}{\zeta^6}
	+ \frac{C_8}{\zeta^8} +
	\frac{C_{10}}{\zeta^{10} } \right ]F(\zeta)  \right \}
\end{equation}
where
$\zeta=r/r_m$.
The function $F(\zeta)$ is given by
$$
 F(\zeta)=\cases { \exp{ \left [ -\left( D/\zeta
 -1 \right) \right]^2},
    & \mbox{if  $\zeta\leq D$}   \cr
1,  & \mbox{if  $\zeta >  D $}\,. }
$$
For completeness the parameters for both HFDHE2 and HFD-B
potentials are given in Table~I.

In the present work  we restrict ourselves to  calculations for
$S$-states only. The partial components $\Phi_{l\lambda 0}$  can
be obtained in this case from the addition of even  partial
waves $l$ and $\lambda$ with  $l=\lambda$.  To  demonstrate the
feasibility of our formalism and the  accuracy which can be
achieved,  we obtained solutions  with $l=0$ and  $l=2$  which
can be compared with other results in the literature.  The
finite--difference approximation in the polar coordinates $\rho$
and  $\theta$ has been used for this purpose, a description of
which is given in the Appendix.

Both potentials  considered, produce a weakly bound state of the
Helium dimer. In our calculations we use the value
$\hbar^2/m=12.12$~K\,\AA$^2$. With this value we found the dimer
energy $\epsilon_d$ was equal to $-0.8301$\,mK in the case of
the HFDHE2 and to $-1.6854$\,mK in the case of the HFD-B
potential. These results are in agreement with other theoretical
results found in the
literature~\cite{Gloeckle,EsryLinGreene,Uang}.  The estimated
experimental value is  $\epsilon_d\sim -1$\,mK
\cite{DimerExp,Science}).  As to the $^4$He atom--$^4$He atom
scattering length, we found that it is 124.7\,{\AA} for the
HFDHE2 and 88.6\,{\AA} for the HFD-B potential.

Since the Helium dimer bound state exists only in the $l=0$
state,  the three-body bound state boundary
conditions~(\ref{BScond}) for the $L=0$  channel reads
\begin{equation}
\label{HeBS}
    \begin{array}{rcl}
 \Phi_{ll0}(x,y) & = & \delta_{l0}\psi_d(x)\exp({\rm i}
 \sqrt{E_t-\epsilon_d}\,y) \left[{\rm a}_0+o\left(y^{-1/2}\right)\right] \\
		& + &
 		\displaystyle\frac{\exp({\rm i}\sqrt{E_t}\rho)}{\sqrt{\rho}}
        	\left[A_{ll 0}(\theta)+o\left(\rho^{-1/2}\right)\right]
\end{array}
\end{equation}
where $E_t$ and $\epsilon_d$ are the trimer and dimer energies
respectively (expressed in units of \AA$^{-2}$)  and
$\psi_d(x)$ stands for the dimer wave function.

The results of the Helium trimer ground-state energy $E_t^{(0)}$
calculations are presented in Table~II. It is seen that they are
in a good agreement with other results  given in the literature.
Although the two  potentials used  differ  only slightly, they
produce  important differences in the ground-state  energy. This
is in agreement with the finding of Ref.~\cite{Sof} but in
disagreement with the statement made in Ref.~\cite{VM4}.  It
should be  further noted that most of the contribution to the
binding energy stems from the  $l=\lambda=0 $  and $l=\lambda=2$
partial component the  latter being more than 35~\%. The
contribution from the $l=\lambda=4 $ partial wave was shown
in~\cite{CGM} to be of the order of a few per cent.

In Ref.~\cite{Gloeckle} Cornelius and Gl\"ockle investigated the
possibility of having Efimov states in the Helium trimer. Their
work  was  motivated by the fact  that the dimer energy
$\epsilon_d$ is very close to the three-body  threshold.
Employing the HFDHE2 potential, these authors found  an  excited
state  at $E_t^{(1)}=-1.6$\,mK.  This finding was recently
confirmed by Esry {\em et al.}~\cite{EsryLinGreene} who also
located an excited state  at $E_t^{(1)}=-1.517$\,mK using the
same HFDHE2 inter-atomic interaction. Note that the approaches
used in~\cite{Gloeckle} and~\cite{EsryLinGreene} are completely
different.  The former  is based on the Faddeev integral
equations in momentum space while  in the later on the
hyperspherical adiabatic approach. In~\cite{EsryLinGreene} the
improved LM2M2 $^4$He--$^4$He potential~\cite{Aziz91} was also
employed and an excited  state at $E_t^{(1)}=-2.118$\,mK was
found.  We have also found that the Helium trimer can form an
excited state with both the HFDHE2 and HFD-B potentials.  The
excited state is present even when only the $l=\lambda=0$
partial wave is taken into account.  This is in agreement with
the finding of Ref.~\cite{Gloeckle}.  Our excited state results
are given in Table~III.  By noting  that the three-body excited
state disappears when the interaction strength increases, both,
Cornelius and Gl\"ockle and Esry {\em et al.}, identify this
state as an Efimov one. We have checked a presence of this
phenomenon in the case of the HFD-B potential and only the
partial wave $l=\lambda=0$ taken into account. Multiplying this
potential by an increasing factor $g\geq1$ we found that
at the beginning the distance
$\delta=\epsilon_d-E_t^{(1)}$, $\delta=\delta(g)$, between
the trimer and dimer energies  $E_t^{(1)}(g)$ and
$\epsilon_d(g)$ increases (see Table~IV) but thereafter (for
$\lambda\ge 1.04$) the $\delta(g)$ monotonically  decreases. 
As can be seen in Table~IV, at  $g\approx1.19$ the value 
of $\delta(g)$ tends to zero, i.e., as 
in~\cite{Gloeckle,EsryLinGreene}, the
excited state disappears (being covered by
the continuous spectrum and probably becoming a
resonance).  This implies the Efimov nature of the 
excited state energy $E_t^{(1)}$. We also performed  calculations 
for a Helium atom scattered  off
a Helium dimer, at $L=0$. For this  we used the asymptotic
boundary conditions (\ref{AsBCPart}) which, for the $L=0$
channel, read
\begin{equation}
\label{AsBCPartS}
    \begin{array}{rcl}
	      	\Phi_{ll0}(x,y,p) & = &
                \delta_{l0}\psi_d(x)\left\{\sin (py) + \exp(py)
        	\left[{\rm a}_0(p)+o\left(y^{-1/2}\right)\right]\right\} \\
		& + &
  \displaystyle\frac{\exp({\rm i}\sqrt{E}\rho)}{\sqrt{\rho}}
        	\left[A_{ll 0}(\theta)+o\left(\rho^{-1/2}\right)\right]
    \end{array}
\end{equation}
The $S$-state elastic scattering phase shifts $\delta_0(p)$ are
then given by
$$
                \delta_0(p)=\frac{1}{2}\,{\rm Im}\,\ln {\rm S}_0(p) 
$$
where ${\rm S}_0(p)=1+2i{\rm a}_0(p)$ is the $(2+1\rightarrow 2+1)$ 
partial component of the scattering matrix. Here, we understand the 
branch of the logarithmic function in such a way that the 
natural ``normalization''  $\delta_0(0)=2\pi$ holds. 

The phase shifts results thus obtained are given in Tables~V, VI.
We considered incident energies below
as well as above the breakup threshold, i.~e., for the
$(2+1\longrightarrow 2+1)$ and the $(2+1\longrightarrow 1+1+1)$
processes. In order to obtain converged results we were
compelled to integrate upto a maximum $\rho_{\rm
max}=400$---600\,\AA.  This comes as no surprise since the
two--body binding energy is very small implying an extended
$^4$He dimer system and thus the trimer wave functions attain
their asymptotic values at very large distances. Changing
$\rho_{\rm max}$ in the range 400---600\,{\AA} produces minimal
effects on the phase shifts. Such a cut-off radius in the three
$^4$He atom problem may be compared with the characteristic
values $\rho_{\rm max}=$20---30\,fm for the three-nucleon
problem (see, for example, the recent paper~\cite{yafkm} and
Refs. therein).  This is not an unexpected result since the
dimer wave function generated, for example,  by the HFD-B
potential, behaves as $\exp(-0.012\,x)$ at large distances (where
$x$ is measured in \AA) while  the deuteron wave function as
$\exp(-0.23\,x)$ (where $x$  is measured in fm).  Thus,  even
scaling considerations imply that 20---30\,fm in the $n-d$
scattering problem are equivalent to 400---600\,{\AA} in the
three $^4{\rm He}$ atoms scattering.

Our estimation for the Helium atom -- Helium dimer scattering
length
$$
\ell_{\rm sc}=-\displaystyle\frac{\sqrt{3}}{2}\,
\begin{array}{c}\phantom{a}\\
{\rm lim}\,\\
\mbox{\scriptsize$p\rightarrow0$}
\end{array}\,
\frac{{\rm a}_0(p)}{p}
$$
with the HFD-B interactions is $170{\pm}5$\,{\AA} in the case if
only the $l=\lambda=0$ partial Faddeev component is taken into
account and $145{\pm}5$\,{\AA} if the two partial waves with
$l=\lambda=0$ and $l=\lambda=2$ are considered. In literature we
found for $\ell_{\rm sc}$ only the result $\ell_{\rm
sc}=195$\,{\AA} of Ref.~\cite{Nakai}, obtained within a
zero-energy scattering calculation based on a separable
approximation for the HFDHE2 potentials.

It is interesting to compare the figures for $\ell_{\rm sc}$
with respective inverse wave numbers $\varkappa^{-1}$ for the
trimer excited state energies. The value of
$\varkappa$ is given by
$
   \varkappa=2\sqrt{(\epsilon_d-E_t^{(1)})/3}
$
where both the $E_t^{(1)}$ and $\epsilon_d$ have to be measured 
in {\AA}$^{-2}$.  Using data of Table~III for the case of the 
HFD-B interaction we find $\varkappa^{-1}\approx 102$\,{\AA} if 
the only partial wave with $l=\lambda=0$ taken into account and 
$\varkappa^{-1}\approx 89$\,{\AA} if the two waves with 
$l=\lambda=0$ and $l=\lambda=2$ are employed. These values are 
about 1.7 times smaller then the values above for the $^4$He 
atom -- $^4$He dimer scattering length found on the base of the 
phase shift results. The situation differs totally from the 
$^4$He two-atomic scattering problem where the inverse wave 
number $\varkappa^{-1}=84.8$\,{\AA} is rather a good 
approximation for the $^4$He--$^4$He scattering length 
$\ell_{\rm sc}^{(2)}=88.6$\,{\AA} mentioned above.  Such a 
significant difference between $\ell_{\rm sc}$ and 
$\varkappa^{-1}$ in the case of the $^4$He three-atomic problem 
may be naturally explained by the Efimov nature of the trimer 
excited state which means that the effective range $r_0$  for 
interaction between $^4$He atom and $^4$He dimer is very large 
as compared to the $^4$He diatomic problem. Unfortunately, 
insufficient accuracy of results for the amplitude $a_0(p)$ at 
$p\approx 0$ which we have at the moment does not allow us to 
extract the values for the $r_0$. 

We present also a number of figures providing a visual 
information about the Helium atom -- Helium dimer scattering.  
The energy $E=+1.4$\,mK situated above the three-body threshold 
has been chosen for this purpose.  In Fig.~2 we plot the Faddeev 
breakup amplitude $A_{ll}(\theta){\equiv}A_{ll0}(\theta)$ for 
$l=0$. The amplitude $A_{22}(\theta)$ is extremely small, 
$|A_{22}(\theta)|^2<5{\cdot}10^{-5}$\,rad$^{-1}$, and therefore 
it is not shown.  The corresponding physical breakup amplitudes 
${\cal A}_{ll}(\theta)\equiv {\cal A}_{ll0}(\theta)$, $l=0,2$, 
are plotted in Fig.~3.  Both figures explicitly exhibit 
the importance of the inclusion of the $l=\lambda=2$ partial 
waves. The large difference between the results obtained for 
the amplitudes ${\cal A}_{l\lambda}(\theta)$ with 
($l=\lambda=0$) and ($l=\lambda=0,2$), implies that higher 
partial waves may also be of importance and should be incorporated 
into the solution in a future research. These findings came as no 
surprise due to the hard--core nature of the underlying 
two--body forces which remain the same for all partial waves. These 
forces generate strong two--body correlations and enhance 
the role played by higher partial waves. 

 As can be also seen in Fig. 3, the breakup is 
rather suppresed in the vicinity of the direction $\theta=30^\circ$ 
while the direction of its maximum lies around $\theta=60^\circ$.  
This behaviour can be understood from the location of the 
domains where particles of a particular pair
are close to each other and thus the elastic 
scattering of the  third particle from the bound pair 
is dominant.

The absolute value of the Faddeev components $\Phi_{ll0}(x,y,p)$ 
for $E=+1.4$\,mK  and $l=\lambda=0$ is plotted in Figs. 4 and 5 
while for   $l=\lambda=2$ in Fig. 6 and 7 
(the corresponding figures for the  partial wave functions 
$\Psi_{ll0}(x,y,p)$ at the same energy 
$E=+1.4$\,mK can be found in Ref.~\cite{MSK-CPL}).  In Fig.~4 
one can explicitly observe the incident ``bound'' plane wave 
$\psi_d(x)\sin(py)$ which is dominant in $\Phi_{000}(x,y,p)$ at large 
distances $y$.  However, such a behaviour is not present in
the partial Faddeev component $\Phi_{220}(x,y,p)$  which differs 
essentially from zero only in the vicinity of the triple collision 
point (see Fig.~6) as  in this domain the coupling between the 
channels $l=\lambda=0$ and $l=\lambda=2$ due to the integral terms of 
Eqs.~(\ref{FTconn}) is large. The rather complicated structure 
of the partial Faddeev components $\Phi_{ll0}(x,y,p)$ in the 
region where the attractions of all  three He--He 
interactions are strong is shown in detail in Figs.~5 and~7.  It 
should be noted that the structure of the functions 
$\Phi_{ll0}(x,y,p)$ (and also of $\Psi_{ll0}(x,y,p)$~\cite{MSK-CPL}) 
practically does not depend on $E$ in the energy range, 
$\epsilon_d<E\lesssim2.4$\,mK, considered 
(compare, for example, our Fig.~5 with Fig.~2  of 
Ref.~\cite{CGM}).

%%%%%%%%%%%%%%%%%%%%%%%%%%%%%%%%%%%%%%%%%%%%%%%%%%%%%%%%%%%%%%%%%%%%
%%%%%%%%%%%%%%%   CONCLUSIONS       %%%%%%%%%%%%%%%%%%%%%%%%%%%%%%
%%%%%%%%%%%%%%%%%%%%%%%%%%%%%%%%%%%%%%%%%%%%%%%%%%%%%%%%%%%%%%%%%%%%

\section{Conclusions}
%%%%%%%%%%%%%%%%%%%%%
In this work we employ a formalism which is suitable for
three--body calculations with hard--core potentials. The
approach is a hard-core variant of the Boundary-Condition  Model
and, unlike some competing methods, is exact and ideally suited
for three-body calculations  with two--body interactions with a
highly repulsive core which  can be treated as a hard--core.
Furthermore the method is feasible not only for bound--state but
for scattering processes as well.

We employed the formalism to calculate  the binding energy of
the $^4$He--trimer system.  The results obtained  with  two
realistic $^4$He--$^4$He potentials  compared favorably with
other results in the literature. Furthermore, we successfully
located an excited state which can be interpreted as an Efimov
state. This clearly demonstrates the reliability of our method
in three-body bound state calculations with hard-core
potentials.  We also endeavored to calculate, for the first
time, the ultra-low energy scattering phase shifts corresponding
to a $^4$He atom scattered off a $^4$He dimer and breakup
amplitudes. Using the phase shift results we gave as well
an estimation for the respective scattering length.

The effectively hard-core  inter-atomic potential together
with other characteristics of the system, make such calculations
extremely tedious and numerically unstable. However, this is not
the case with our approach where the hard--core  is taken from
the beginning into account in a mathematically rigorous way. The
successful application of the proposed method revealed that this
method is ideally  suited for  calculations in systems where the
strong repulsion in the pairwise forces gives rise to strong
numerical inaccuracies which  make  calculations for these
molecules cumbersome. Thus the formalism  paves the way  to
study various ultra--cold three-atomic systems, and to calculate
important quantities such as the cross-sections,
recombination rates {\it etc.}

%%%%%%%%%%%%%%%%%%%%%%%%%%%%%%%%%%%%%%%%%%%%%%%%%%%%%%%%%%%%%%%%%%%%
%%%%%%%%%%%%%%%   ACKNOWLEDGEMENTS  %%%%%%%%%%%%%%%%%%%%%%%%%%%%%%
%%%%%%%%%%%%%%%%%%%%%%%%%%%%%%%%%%%%%%%%%%%%%%%%%%%%%%%%%%%%%%%%%%%%

\bigskip
\acknowledgements
Financial support from the University of South Africa, the Joint
Institute for Nuclear Research, Dubna, and the Russian
Foundation for Basic Research (Projects No.~96-01-01292,
No.~96-01-01716 and No.~96-02-17021) is  gratefully
acknowledged.  The authors are indebted to Dr.~F.~M.~Penkov for
a number of useful remarks and to Prof.~I.~E.~Lagaris for
allowing us to use the computer facilities of the University of
Ioannina, Greece, to perform the excited state and
scattering  calculations.

%%%%%%%%%%%%%%%%%%%%%%%%%%%%%%%%%%%%%%%%%%%%%%%%%%%%%%%%%%%%%%%%%%%%
%%%%%%%%%%%%%%%     APPENDIX    %%%%%%%%%%%%%%%%%%%%%%%%%%%%%%%%%%%%
%%%%%%%%%%%%%%%%%%%%%%%%%%%%%%%%%%%%%%%%%%%%%%%%%%%%%%%%%%%%%%%%%%%%
\newpage
\appendix
\section*{Numerical method}

The finite-difference approximation in polar coordinates $\rho$
and $\theta$ has been used to solve this problem.  For this, the
grid knots were  chosen to be the points of intersection of the
arcs $\rho=\rho_i$, $i=1,2,\ldots, N_\rho$, and the rays
$\theta=\theta_j$, $j=1,2,\ldots, N_\theta$.  The  value of the
parameter $c$ (``diameter'' of the particle cores) is chosen in
such a way that any further decrease of it does not affect the
trimer ground-state energy. In the present case a four figure
accuracy  has been achieved with $c=0.7$\,{\AA}.

The $\rho_i$ points were chosen according to the formulas
\begin{eqnarray}
\nonumber
\rho_i &=&\frac{i}{N_c^{(\rho)}+1}\, c, \quad i=1,2,\ldots,N_c^{(\rho)},\\
\nonumber
		\rho_{i+N_c^{(\rho)}}& = & \sqrt{c^2 + y_i^2}, \quad
	     	i=1,2,\ldots,N_\rho-N_c^{(\rho)},
\end{eqnarray}
where $N_c^{(\rho)}$ stands for the number of arcs inside of the
core domain and
$$
		y_i = f(\tau_i)\sqrt{\rho^2_{N_\rho}-c^2}, \quad
		\tau_i = \frac{i}{N_\rho-N_c^{(\rho)}}.
$$
The non-linear monotonously increasing function $f(\tau)$,
$0\leq\tau\leq 1$, satisfying the conditions $f(0)=0$ and
$f(1)=1$ was chosen in the form
$$
  f(\tau)=\displaystyle\frac{(1+{\sf a})\tau^2}{1+ {\sf a}\tau}
$$
in the case of the ground-state calculations and in the form
$$
		f(\tau)=\left\{
		\begin{array}{lcl} \alpha_0\tau & , & \tau\in[0,\tau_0]\\
		\alpha_1\tau+\tau^\nu &,& \tau\in(\tau_0,1]
		\end{array}
		\right..
$$
in the case of scattering and excited state calculations. A
typical value of the ``acceleration'' ${\sf a}$, ${\sf a}\geq
0$, which is  satisfactory in  ground-state calculations is
${\sf a}=0.4$ (for $\rho_{N_\rho}< 100$\,{\AA}). The values of
$\alpha_0$, $\alpha_0\geq 0,$ and $\alpha_1$, $\alpha_1\geq 0,$
are defined via $\tau_0$ and $\nu$ from the continuity condition
for $f(\tau)$ and its derivative at the point $\tau_0$. A
typical value of $\tau_0$ is $0.2$. The value of the power $\nu$
 depends on the cut-off radius $\rho_{N_\rho}=$200---600\,{\AA}
its  range being within 3.3---4.75.

The knots $\theta_j$ for $j=1,2,\ldots,N_\rho-N_c^{(\rho)}$ were 
taken according to $\theta_j=\arctan(y_j/c)$.  The rest knots 
$\theta_j$, $j=N_\rho-N_c^{(\rho)}+1, \ldots,N_\theta,$ were 
chosen equidistantly.  Such a choice of the grid is prescribed 
by the need to have the points of intersection of the arcs 
$\rho=\rho_i$ and the rays $\theta=\theta_j$ with the core line 
$x=c$ as its knots.  Furthermore, the grid must be constructed 
in such a manner so that the density of the points is higher 
where the Faddeev components are important, i.~e., for small  
values of  $\rho$ and/or $x$, and lower in the asymptotic 
region. Usually we took the same numbers of grid points for both 
$\theta$ and $\rho$, $N_\theta=N_\rho$. For  $N_c$ we chose 
$N_c^{(\rho)}=5$.

The maximal $\rho$ value used, $\rho_{\rm max}=\rho_{N_\rho}$,
in our ground-state of the Helium $^4$He trimer calculations was
60\,{\AA}.  Beyond this radius the effects on the bound state
are minimal.  For the excited state calculations we were obliged
to increase the $\rho_{\rm max}$ to 200---400\,{\AA}  and for
the scattering calculations to 400---600\,{\AA}.

A description of a finite-difference algorithm of solving the
Faddeev differential equations for conventional potentials was
given in~\cite{MF,MGL}. A generalization of this algorithm to
the boundary-condition model for the three-nucleon problem was
previously employed
in~\cite{MotovilovVLGU,MotovilovThesis,EChAYa}.  Here, we shall
describe in more detail an extension of the algorithm \cite{MGL}
to  the hard-core boundary conditions problems.  For simplicity
we restrict ourselves to the $(2+1\rightarrow2+1\,;\,1+1+1)$
scattering and the bound-state boundary-value problems where
only one Faddeev partial equation with $l=0$ is considered.

In the scattering problem, we firstly, in
the component $\Phi(x,y,p)\equiv \Phi_{000}(x,y,p)$
explicitly separate
the initial-state wave function $\chi(x,y,p)=\psi_d(x)\sin(py)$
(see~(\ref{AsBCPartS})). As a result, (\ref{FadPart})
and~(\ref{BCCorePart})  are reduced to inhomogeneous equations
for the remainder $\Phi'=\Phi-\chi$ which differ in form  from
(\ref{FadPart}) and~(\ref{BCCorePart}) only by the presence on
the right-hand side of inhomogeneous terms 
$F^r(x,y)$ and  $F^c(y)$, respectively, whose explicit form 
is obvious. 

On a fixed arc $\rho=\rho_i$ of the polar grid concerned, the
values of the function $\Phi'$ and inhomogeneous
terms $F^r(x,y)$ and $F^c(y)$ form vectors ${\cal X}^{(i)}
\in{{\Bbb C}}^{N_\theta}$, ${\cal F}^{(i)}\in{{\Bbb R}}^{N_\theta}$, having
components ${\cal X}^{(i)}_j
=\Phi'(\rho_i\cos\theta_j,\rho_i\sin\theta_j)$  and ${\cal
F}^{(i)}_j=F^r(\rho_i\cos\theta_j, \rho_i\sin\theta_j)$ if
$\rho_i\cos\theta_j\neq c$ or ${\cal
F}^{(i)}_j=F^c(\rho_i\sin\theta_j)$ if $\rho_i\cos\theta_j=
c$. The set of vectors ${\cal X}^{(i)}$, \, ${\cal F}^{(i)}$,
$i=1,2,\ldots,N_\rho$,  determines the vectors ${\cal
X}\in{{\Bbb C}}^{N_{\theta\rho}}$ and ${\cal
F}\in{{\Bbb R}}^{N_{\theta\rho}}$, $N_{\theta\rho}=N_\theta N_\rho$:
$     
     {\cal X}=\begin{array}{c}
     \mbox{\scriptsize$N_\rho$}\\[-3pt]
     \oplus\\[-3pt] \mbox{\scriptsize$i=1$}
    \end{array}{\cal X}^{(i)}, \quad
    {\cal F}=\begin{array}{c}
     \mbox{\scriptsize$N_\rho$}\\[-3pt]
     \oplus\\[-3pt] \mbox{\scriptsize$i=1$}
    \end{array}{\cal F}^{(i)}.
$
In such a representation, Eqs.~(\ref{FadPart})
and(\ref{BCCorePart}) assumed the form
\begin{equation}
\label{FinDif}
	  \left\{
	\begin{array}{l} {\cal X}^{(0)}=0\,, \\
		L_i{\cal X}^{(i-1)}+(M_i-E\tilde{I}_i)
		{\cal X}^{(i)}+R_i{\cal X}^{(i+1)}
		={\cal F}^{(i)}, \quad i=1,2,\ldots,N_\rho\,.
	\end{array}
	  \right.
\end{equation}
Here, $L_i,$ $M_i,$ $\tilde{I}_i$ and $R_i$ are matrices of rank
$N_\theta$. The matrices $L_i$ and $R_i$ are generated only by
the radial part of the Laplacian in (\ref{FadPart}) and are
therefore diagonal. The non-diagonal matrix $M_i$ describes the
contribution of the central terms of the radial part of the
Laplacian, of its spherical part, the potential, and the
integral operator on the arc $\rho=\rho_i\,.$ In the cases where
$i,j$ are such that $\rho_i \cos\theta_j=c$, the respective rows
of the matrices $L_i,$ $M_i,$ and $R_i$ are generated by the
condition~(\ref{BCCorePart}). The matrix $\tilde{I}_i$ differs
from the unity one only in a row corresponding to the boundary
condition~(\ref{BCCorePart}).  This row in $\tilde{I}_i$ has
zero elements.

The system~(\ref{FinDif})  includes $N_{\theta\rho}$ equations
for $N_{\theta\rho}+N_\theta$ unknowns. An additional relation
that selects a unique solution of (\ref{FinDif}) follows from
the asymptotic conditions~(\ref{AsBCPartS}):
\begin{equation}
\label{MCond}
   {\cal X}^{(N_ \rho+1)}=B_{N_\rho}\tilde{I}_{N_\rho}{\cal X}^{(N_\rho)}+
   {\rm a}_0(p)\tilde{I}_{N_\rho}{\cal D}^{(N_\rho)}
\end{equation}
where $B_{N_\rho}={\rm
diag}\{b_1,b_2,\ldots,b_{N_{\theta}}\}$ is a diagonal matrix
with elements
$$
        b_j=C^{(+)}_{N_\rho}\left[1+o(\rho_{N_\rho}^{-1/2})\right]\,, \quad
  C^{(+)}_{N_\rho}=\sqrt{\displaystyle\frac{\rho_{N_\rho}}{\rho_{N_\rho+1}}}
         \exp[{\rm i}\sqrt{E}(\rho_{N_\rho+1}-\rho_{N_\rho})]\,,
$$
and ${\cal{D}}^{(N_\rho)}$,
${\cal{D}}^{(N_\rho)}\in{{\Bbb C}}^{N_\theta}$, is a vector with
components
$
     {\cal{D}}^{(N_\rho)}_j=\chi_1(\rho_{N_\rho+1},\theta_j)
      -b_j\chi_1(\rho_{N_\rho},\theta_j)
$
where
$
\chi_1(\rho,\theta)=
\psi_d(\rho\cos\theta)  \exp({\rm i}\,p\,\rho\sin\theta).
$

The condition~(\ref{MCond}) allows the elimination of  ${\cal
X}^{(N_\rho+1)}$ and reduces the last equation  of the
system~(\ref{FinDif}) to
\begin{equation}
\label{ENR}
L_{N_\rho}{\cal X}^{(N_\rho-1)}+ (\tilde{M}_{N_\rho}-E\tilde{I}_{N_\rho})
          {\cal X}^{({N_\rho})}={\cal F}^{({N_\rho})}
         +{\rm a}_0(p)\tilde{\cal F}^{({N_\rho})}
\end{equation}
where the matrix $\tilde{M}_{N_\rho}$ and the vector
$\tilde{\cal F}^{(N_\rho)}$
 are given by
$\tilde{M}_{N_\rho}=M_{N_\rho}+R_{N_\rho}B_{N_\rho}\tilde{I}_{N_\rho}$
and
$\tilde{\cal F}^{(N_\rho)}=
  R_{N_\rho}\tilde{I}_{N_\rho}{\cal D}^{(N_\rho)}\,.$

The system (\ref{FinDif}), after replacing  its last equation
with (\ref{ENR}), can be written in the form
\begin{equation}
\label{MFinal}
		(K-E\tilde{I}){\cal X}={\cal F}+{\rm a}_0(p){\cal F}'
\end{equation}
where $K$ is a three-block-diagonal matrix constructed of the
blocks $L_i$, $M_i$ (or $\tilde{M}_{N_\rho}$ if $i=N_\rho$), and
$R_i$, $i=1,2,\ldots,N_\rho$, while $\tilde{I}$,
$\tilde{I}=\begin{array}{c}
     \mbox{\scriptsize$N_\rho$}\\[-3pt]
     \oplus\\[-3pt] \mbox{\scriptsize$i=1$}
    \end{array}\tilde{I}_i\,$,
is a diagonal matrix. Both $K$ and $\tilde{I}$ are matrices of
rank $N_{\theta\rho}$. From (\ref{FinDif}), it follows that $K$
has a band structure with band width $2N_\theta+2$. The vector
${\cal F}'$ in~(\ref{MFinal}) reads as ${\cal
F}'=\begin{array}{c}
     \mbox{\scriptsize$N_\rho$}\\[-3pt]
     \oplus\\[-3pt] \mbox{\scriptsize$i=1$}
    \end{array}{{\cal F}'}^{(i)}$ with
${{\cal F}'}^{(i)}=0,$ $i=1,2,\ldots,N_\rho-1,$ and ${{\cal
F}'}^{(N_\rho)}= \tilde{\cal F}^{(N_\rho)}.$

The solution of~(\ref{MFinal}) can be expressed as
\begin{equation}
\label{SolSepar}
		{\cal X}={\cal X}_0+{\rm a}_0(p){\cal X}_1
\end{equation}
where the vectors ${\cal X}_0$ and ${\cal X}_1$ are determined from
\begin{equation}
\label{Separ}
(K-E\tilde{I}){\cal X}_0={\cal F}\,;
\quad (K-E\tilde{I}){\cal X}_1={\cal F}'
\end{equation}
in which the inhomogeneous terms are known.

Having determined the vectors ${\cal X}_0$ and ${\cal X}_1$, we
can then proceed, via the asymptotics~(\ref{AsBCPartS}), to find
the elastic scattering amplitude ${\rm a}_0(p)$.  For this we
may use two methods. In the first one, we compare the
representations (\ref{AsBCPartS}) and~(\ref{SolSepar}) on the
arc $\rho=\rho_{N_\rho}$ in those knots
$(\rho_{N_\rho},\theta_j)$ of the grid for which the value of
$\rho_{N_\rho}\cos\theta_j$ belongs to a vicinity of the point
$x_0$ where the dimer wave function $\psi_d(x)$ is maximal,
$\psi_d(x_0)={\rm max}\,\psi_d(x)$. In this vicinity,
the term with the spherical wave ${\exp({\rm
i}\sqrt{E}\rho)}/{\sqrt{\rho}}$ is much smaller than the
``elastic'' wave term $\psi_d(x)\exp({\rm i}py)$ (if
$\rho_{N_\rho}$ is sufficiently large). Therefore, omitting the
breakup term we obtain from (\ref{SolSepar}) the following
expression
\begin{equation}
\label{AsM1}
  {\rm a}_0(p)=\frac{\left[{\cal X}_0^{(N_\rho)}\right]_j}
  {\chi_1(N_\rho,\theta_j)-\left[{\cal X}_1^{(N_\rho)}\right]_j }
\end{equation}
where the index $j$ corresponds to the angles $\theta_j$ for
which $\rho_{N_\rho}\cos\theta_j\approx x_0.$

In the second method we compare the components of
(\ref{SolSepar}) with the asymptotic representation
(\ref{AsBCPartS}) on  two successive arcs $\rho=\rho_{N_\rho-1}$
and $\rho=\rho_{N_\rho}$, omitting only the terms
$\psi_d(x)o(y^{-1/2})$ and
$\exp({\rm i}\sqrt{E}\rho)o(\rho^{-1})$.
As a result we find
\begin{equation}
\label{AsM2}
    {\rm a}_0(p)=-\displaystyle\frac{\left[{\cal X}_0^{(N_\rho)}\right]_j-
    C^{(-)}_{N_\rho}\left[{\cal X}_0^{(N_\rho-1)}\right]_j}
    {\left[{\cal X}_1^{(N_\rho)}\right]_j-\chi_1(\rho_{N_\rho},\theta_j)-
    C^{(-)}_{N_\rho}  \left\{\left[{\cal X}_1^{(N_\rho-1)}\right]_j-
    \chi_1(\rho_{N_\rho-1},\theta_j)\right\} }
\end{equation}
with
$
   C^{(-)}_{N_\rho}=\sqrt{\displaystyle\frac{\rho_{N_\rho-1}}{\rho_{N_\rho}}}
   \exp[{\rm i}\sqrt{E}(\rho_{N_\rho}-\rho_{N_\rho-1})].
$
As in (\ref{AsM1}), the index $j$ corresponds to a vicinity of
the point $x_0$ where the dimer wave function acquires a
maximal value.

Having calculated ${\rm a}_0(p)$ via~(\ref{AsM1}) or
(\ref{AsM2}) we can find, using (\ref{SolSepar}), the vector
${\cal X}^{(N_\rho)}$  corresponding to the values of the
desired function $\Phi'$ on the final arc $\rho=\rho_{N_\rho}$,
$\Phi'(\rho_{N_\rho}\cos\theta_j, \rho_{N_\rho}\sin\theta_j)=
{\cal X}^{(N_\rho)}_j$, and then determine the Faddeev
breakup amplitude
$$
     A_{000}(\theta_j)=\left[
   {\cal X}^{(N_\rho)}_j-{\rm a}_0(p)
   \chi_1(\rho_{N_\rho},\theta_j) \right]
   \sqrt{\rho_{N_\rho}}\exp(-{\rm i}\sqrt{E}\rho_{N_\rho})\,.
$$

In the bound-state problem we deal with the same system of
equations (\ref{FinDif}) for ${\cal
X}^{(i)}_j=\Phi(\rho_i\cos\theta_j,
\rho_i\sin\theta_j)$ where now $\Phi(x,y)$ stands for a
bound-state wave function satisfying the asymptotic conditions
(\ref{HeBS}). Of course the inhomogeneous terms ${\cal
F}^{(i)}$ vanish in this case.

To eliminate the vector ${\cal X}^{(N_\rho+1)}$ from the last
$(i=N_\rho)$ equation of (\ref{FinDif}) we use the
representation (\ref{HeBS}). For angles corresponding to the
knots of the arc $\rho=\rho_{N_\rho}$ lying inside the core
domain, $\rho_{N_\rho}\cos \theta_j<c$, we write the components
${\cal X}^{(N_\rho)}_j$ and ${\cal X}^{(N_\rho+1)}_j$ on the two
successive arcs $\rho=\rho_{N_\rho}$ and $\rho=\rho_{N_\rho+1}$,
taking into account the condition $\psi_d(x)=0$, $x\leq c$,
and neglecting the terms $\exp({\rm
i}\sqrt{E}\rho)o(\rho^{-1})$. Then we find
$
    {\cal X}_j^{(N_\rho+1)}=C_{N_\rho}^{(+)}{\cal X}_j^{(N_\rho)}\,.
$
For angles $\theta_j$ corresponding to  knots of the arc 
$\rho=\rho_{N_\rho}$ lying outside the core domain, 
$\rho_{N_\rho}\cos\theta_j>c$, we write the 
expression~(\ref{HeBS}) for the components ${\cal X}_j^{(i)}$ on 
three successive arcs $\rho=\rho_i$, 
$i=N_{\rho}-1,N_{\rho},N_{\rho}+1,$ neglecting the terms 
$\psi_d(x)o(y^{-1})$ and $\exp({\rm 
i}\sqrt{E}\rho)o(\rho^{-1})$.
Using this expression for $i=N_\rho-1$ and $i=N_\rho$ we can
express ${\rm a}_0$ and $A(\theta_j)$ in terms of ${\cal
X}_j^{(N_\rho-1)}$ and ${\cal X}_j^{(N_\rho)}$. Then we may
express  ${\cal X}_j^{(N_\rho+1)}$ in terms of ${\cal
X}_j^{(N_\rho-1)}$ and ${\cal X}_j^{(N_\rho)}$ 
using~(\ref{HeBS})   for $i=N_{\rho}+1$. Thus, finally, the last
equation of (\ref{FinDif}) becomes
$$
 \tilde{L}_{N_\rho}{\cal X}^{(N_\rho-1)}+
 (\tilde{M}_{N_\rho}-E\tilde{I}_{N_\rho})
 {\cal X}^{({N_\rho})}=0
$$
where the  matrices $\tilde{L}_{N_\rho}$ and
$\tilde{M}_{N_\rho}$  are given by
\begin{equation}
\label{Elim}
   \tilde{L}_{N_\rho}=L_{N_\rho}+
   R_{N_\rho}\tilde{I}_{N_\rho}{W}_{N_\rho}\,,
   \quad \tilde{M}_{N_\rho}=M_{N_\rho}+ R_{N_\rho}
   \tilde{I}_{N_\rho}\tilde{W}_{N_\rho}\,.
\end{equation}
The $W={\rm diag}\{w_1^{(N_\rho)},\ldots,
w_{N_\theta}^{(N_\rho)}\}$ and $\tilde{W}={\rm diag} \{
\tilde{w}_1^{(N_\rho)},\ldots, \tilde{w}_{N_\theta}^{(N_\rho)}
\} $ are diagonal matrices with
\begin{equation}
\label{W}
   w_j^{(N_\rho)}=
   \left\{
   \begin{array}
   {ccc} \displaystyle\frac{
   \chi_1(\rho_{N_\rho+1},\theta_j)-C^{(+)}_{N_\rho} \chi_1(\rho_{N_\rho},
   \theta_j) }
   { \chi_1(\rho_{N_\rho-1},\theta_j)-{C^{(-)}_{N_\rho}}^{-1}
\chi_1(\rho_{N_\rho},\theta_j) } & , & \rho_{N_\rho}\cos\theta_j >c\,,\\
    0\mbox{\phantom{\Large I}}&,& \rho_{N_\rho}\cos\theta_j \leq c \,,
    \end{array}
   \right.
\end{equation}
and
\begin{equation}
\label{Wt}
  \tilde{w}_j^{(N_\rho)}=\left\{
  \begin{array}{ccc}  \displaystyle
  \frac {C^{(+)}_{N_\rho}\chi_1(\rho_{N_\rho-1},\theta_j)-
  {C^{(-)}_{N_\rho}}^{-1}
  \chi_1(\rho_{N_\rho+1},\theta_j)} {\chi_1(\rho_{N_\rho-1},\theta_j)
  -{C^{(-)}_{N_\rho}}^{-1}\chi_1(\rho_{N_\rho},\theta_j)}
  &,& \rho_{N_\rho}\cos\theta_j >c\,,\\
  C^{(+)}_{N_\rho}\mbox{\phantom{\Large I}} &,& \rho_{N_\rho}
  \cos\theta_j \leq c\,,
  \end{array}\right.
\end{equation}
where now
$
    \chi_1(\rho,\theta)= \psi_d(\rho\cos\theta)\exp({\rm i}
    \sqrt{E-\epsilon_d}\,\rho\sin\theta )
$.
Note that the matrices $\tilde{L}_{N_\rho}$ and
$\tilde{M}_{N_\rho}$ depend on the energy $E$ since the function
$\chi_1$ and the coefficients $C^{(\pm)}_{N_\rho}$ are functions of
$E$.  Therefore the total matrix $K$ of the system obtained is
also a function of $E$, $K=K(E)$. In this work we searched for
binding energies of the $^4$He trimer  as roots of the
determinant ${\rm det}\{K(E)-E\tilde{I}\}$.

The use of the asymptotic boundary conditions~(\ref{HeBS}) in
the form of (\ref{Elim})--(\ref{Wt}) is extremely important when
searching for the excited $E_t^{(1)}$ state. It is difficult to
locate this state if the term ${\rm a}_0\psi_d(x)\exp({\rm
i}\sqrt{E-\epsilon_d}\,y)$ is omitted.  This means that the
dimer wave function $\psi_d(x)$ gives a decisive contribution
into the excited state. In contrast, omitting this term (and
replacing $w_j^{(N_\rho)}$, $\tilde{w}_j^{(N_\rho)}$ given
by~(\ref{W}) and (\ref{Wt}) with $w_j^{(N_\rho)}=0,$
$\tilde{w}_j^{(N_\rho)}=C^{(+)}_{N_\rho}$ for all
$j=1,\ldots,N_\theta$) in the ground-state calculations
simplifies the problem considerably by allowing the decrease of
the cut-off radius to 60\,{\AA}. Otherwise, to obtain the
correct result for $E_t^{(0)}$  we had to increase
$\rho_{N_\rho}$ up to 150---200\,{\AA} as  the dimer wave
function falls off slowly.

%%%%%%%%%%%%%%%%%%%%%%%%%%%%%%%%%%%%%%%%%%%%%%%%%%%%%%%%%%%%%%%%%%%%
%%%%%%%%%%%%%%%   REFERENCES  %%%%%%%%%%%%%%%%%%%%%%%%%%%%%%%%%%%%%%
%%%%%%%%%%%%%%%%%%%%%%%%%%%%%%%%%%%%%%%%%%%%%%%%%%%%%%%%%%%%%%%%%%%%

%%%%%%%%%%%%%%%%%%%%%%%%%%%%%%%%%%%%%%%%%%%%%%%%%%%%%%%%%%%
%%%%%%%%%%%%%%%   TABLES     %%%%%%%%%%%%%%%%%%%%%%%%%%%%%%
%%%%%%%%%%%%%%%%%%%%%%%%%%%%%%%%%%%%%%%%%%%%%%%%%%%%%%%%%%%

%%%%%%%%%%%%%%%   TABLE  I    %%%%%%%%%%%%%%%%%%%%%%%%%%%%%%

\begin{table}
\label{tableI}
\caption{The parameters for the $^4$He$-$$^4$He
          potentials used. }
\begin{tabular}{ccl}
\hline
Parameter & HFDHE2 \cite{Aziz79} & HFD-B \cite{Aziz87}  \\\\\hline
   $\varepsilon$ (K)      &    10.8       &  10.948     \\
   $ r_m $ (\AA)          &   2.9673      &  2.963      \\
   $A$                    &   544850.4    &  184431.01  \\
   $\alpha$               & 13.353384     & 10.43329537 \\
   $\beta$                &     0         & $-2.27965105$ \\
   $C_6$                  & 1.3732412     & 1.36745214  \\
   $C_8$                  & 0.4253785     & 0.42123807  \\
   $C_{10}$               & 0.178100      & 0.17473318  \\
   $D$                    & 1.241314      & 1.4826\\\\
\hline
\end{tabular}
\end{table}
%%%%%%%%%%%%%%%   TABLE  II    %%%%%%%%%%%%%%%%%%%%%%%%%%%%%%

\begin{table}
\label{tableII}
\caption
{Bound state energy $E_t^{(0)}$ results for the Helium trimer.
The (absolute) values of $E_t^{(0)}$ are given in K.
The grid parameters used were: $N_\theta=N_\rho=275$, ${\sf a}=0.4$,
and $\rho_{\rm max}=$60\,{\AA}.}
\begin{tabular}{|c|ccccc|cc|c|}
\hline
Potential & \multicolumn{5}{c|}{Faddeev equations}
&\multicolumn{2}{c|}{Variational}&\multicolumn{1}{c|}{Adiabatic} \\
 & \multicolumn{5}{c|}{}
&\multicolumn{2}{c|}{methods}&\multicolumn{1}{c|}{approach} \\
\cline{2-9}
   & $l$ &  This work & \cite{CGM} & \cite{Gloeckle}
& \cite{Nakai} & \cite{VMC2} & \cite{VM4}
& \multicolumn{1}{c|}{\cite{EsryLinGreene}}\\
%   &     &         & & & & &\multicolumn{1}{c|}{} \\
\hline \hline
HFDHE2 & 0 & $0.084$ &  & $0.082$ & $0.092$ &  &  & $0.098$
\\ \cline{2-6}
 & 0,2 &  $0.114$ & $0.107$ & $0.11$ &  & $0.1173$ &  &  \\
\hline \hline
 HFD-B & 0 &$0.096$ & $0.096$ & & & & & \\
\cline{2-6}
 & 0,2 & $0.131$  & $0.130$ & &  & & $0.1193$ & \\
\hline
\end{tabular}
\end{table}

%%%%%%%%%%%%%%%   TABLE  III    %%%%%%%%%%%%%%%%%%%%%%%%%%%%%%

\begin{table}
\label{tableIII}
\caption
{Excited state energy $E_t^{(1)}$ results for the Helium trimer.
The (absolute) values of $E_t^{(1)}$ are given in mK.
The grid parameters used were: $N_\theta=N_\rho=252$, $\tau_0=0.2$,
$\nu=3.6$ and $\rho_{\rm max}=$250\,{\AA}.}
\begin{tabular}{|c|c|c|c|c|c|}
\hline
Potential   & $l$ &  This work &  \cite{Gloeckle} & \cite{Nakai}
					  &\cite{EsryLinGreene}\\
\hline \hline
HFDHE2      &  0  &  $1.5$     &  $1.46$    &  $1.04$ & $1.517$  \\
\cline{2-5}
            & 0,2 &  $1.7$     &   $1.6$     &   & \\
\hline \hline
 HFD-B      &  0  &  $2.5$     &               &  &\\
\cline{2-5}
            & 0,2 &  $2.8$     &               &  &\\
\hline
\end{tabular}
\end{table}

%%%%%%%%%%%%%%%   TABLE  IV    %%%%%%%%%%%%%%%%%%%%%%%%%%%%%%

\begin{table}
\label{tableIV}
\caption
{Dependence of the dimer bound state, $\epsilon_d$, and trimer
excited state, $E_t^{(1)}$, energies on the multiplying factor
(potential strength) $g$ in the case of the HFD-B $^4$He--$^4$He
interaction.  The values of energies $\epsilon_d$, $E_t^{(1)}$
and their difference $\delta=\epsilon_d-E_t^{(1)}$ are given in
mK.  The grid parameters used were:  $N_\theta=N_\rho=550$,
$\tau_0=0.215$, $\nu=4.75$ and $\rho_{\rm max}=$350\,{\AA}.}
\begin{tabular}{|c|c|c|c|}
\hline
$g$   & $\epsilon_d$ &  $E_t^{(1)}$  & $\delta$ \\
\hline
0.975  & $  -0.036 $  & $ -0.308 $ &  0.272 \\
 1.00  & $  -1.685 $  & $ -2.485 $ &  0.800 \\
 1.04  & $  -9.368 $  & $-10.353 $ &  0.985 \\
 1.10  & $ -32.222 $  & $-32.777 $ &  0.556 \\
 1.16  & $ -68.150 $  & $-68.334 $ &  0.184 \\
\hline
\end{tabular}
\end{table}

%%%%%%%%%%%%%%%   TABLE  V    %%%%%%%%%%%%%%%%%%%%%%%%%%%%%%

\begin{table}
\label{tableV}
\caption
{Phase shift results for the $L=0$, $l=\lambda=0$ partial wave
obtained with the HFD-B $^4$He--$^4$He potential.  The grid
parameters used were:  $N_\theta=N_\rho=320$, $\tau_0=0.2$,
$\nu=4.5$, and $\rho_{\rm max}=$400\,{\AA}.  }
\begin{tabular}{|cc|cc|cc|}
\hline
 $E$ (mK)  & $\delta_0$ (deg) &  $E$ (mK)  & $\delta_0$ (deg) &
 $E$ (mK)  & $\delta_0$ (deg) \\
\hline
$-1.68535$  &  359.2  & $ -1.4  $   &  315.8  &    0.1  & 276.8 \\
$-1.6853 $  &  358.9  & $ -1.25 $   &  309.0  &    0.4  & 272.6 \\
$-1.685 $  &  357.5  & $ -1.1  $   &  303.6  &    0.7  & 268.9 \\
$-1.68  $  &  352.3  & $ -0.95 $   &  299.1  &    1.0  & 265.7 \\
$-1.67  $  &  347.2  & $ -0.8  $   &  295.0  &    1.4  & 261.8 \\
$-1.66  $  &  344.2  & $ -0.7  $   &  292.5  &    1.6  & 260.0 \\
$-1.65  $  &  341.4  & $ -0.5  $   &  287.9  &    1.8  & 258.4 \\
$-1.60  $  &  333.4  & $ -0.4  $   &  285.7  &    2.1  & 256.2 \\
$-1.5   $  &  322.3  & $ -0.2  $   &  281.7  &    2.4  & 254.2 \\
\hline
\end{tabular}
\end{table}

%%%%%%%%%%%%%%%   TABLE  VI    %%%%%%%%%%%%%%%%%%%%%%%%%%%%%%

\begin{table}
\label{tableVI}
\caption {As in Table~V but with the inclusion of the
$l=\lambda=2$ partial wave. }
\begin{tabular}{|cc|cc|cc|}
\hline
 $E$ (mK)  & $\delta_0$ (deg) &  $E$ (mK)  & $\delta_0$ (deg) &
 $E$ (mK)  & $\delta_0$ (deg) \\
\hline
$-1.68535$ & 359.3   & $-1.4 $   &  323.0   &  0.3   &   283.7   \\
$-1.6853$ &  359.0   & $-1.1 $   &  312.6   &  0.7   &   278.7   \\
$-1.685 $ &  357.8   & $-0.8$   &   304.6   &  1.0   &   275.4   \\
$-1.68$   &  353.3   & $-0.6$   &   299.8   &  1.4   &   271.5   \\
$-1.60$   &  336.4   & $-0.4$   &   295.5   &  1.8   &   268.0   \\
$-1.5 $   &  328.0   & $-0.1$   &   289.7   &  2.4   &   263.5   \\
\hline
\end{tabular}
\end{table}

%%%%%%%%%%%%%%%%%%%%%%%%%%%%%%%%%%%%%%%%%%%%%%%%%%%%%%%%%%%
%%%%%%%%%%%%%%%   FIGURES    %%%%%%%%%%%%%%%%%%%%%%%%%%%%%%
%%%%%%%%%%%%%%%%%%%%%%%%%%%%%%%%%%%%%%%%%%%%%%%%%%%%%%%%%%%
\bigskip

\bigskip
%%%%%%%%%%%%%%%   FIGURE  1  %%%%%%%%%%%%%%%%%%%%%%%%%%%%%%

\begin{figure}
\epsfig{file=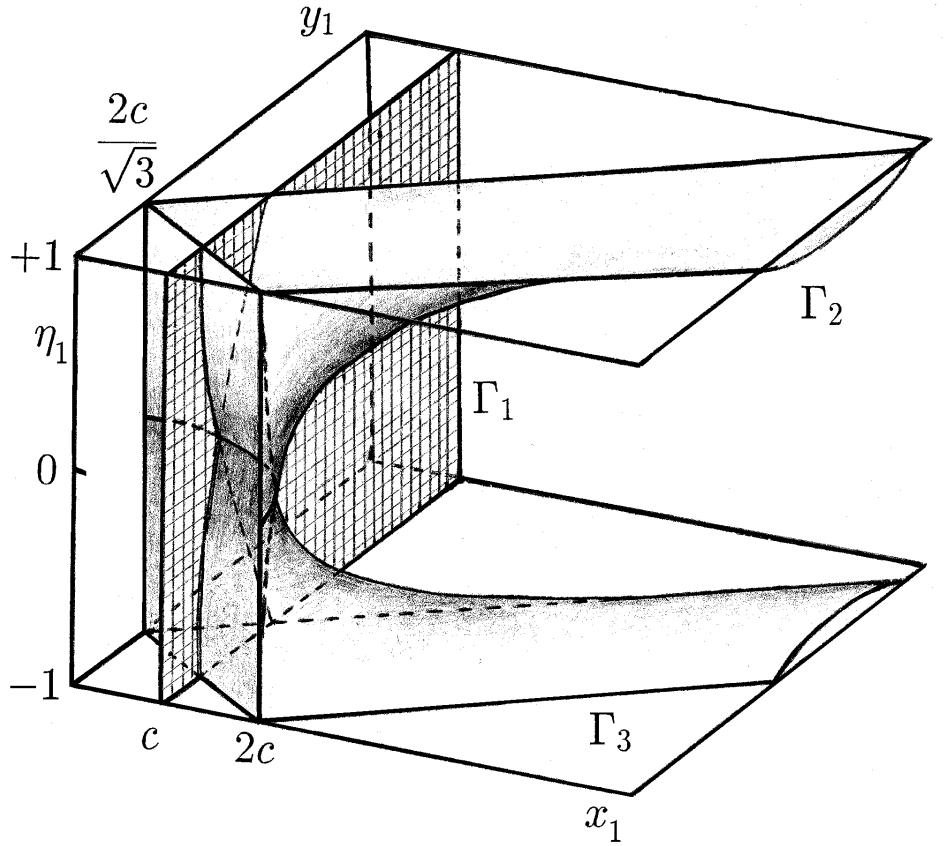,height=12cm}
\caption{Three-dimensional image of the three-body configuration
space for particles with equal masses and the same core radii.
See explanation of the notations used in Sect.~II.}
\label{Fig-cs}
\end{figure}

%%%%%%%%%%%%%%%   FIGURE  2  %%%%%%%%%%%%%%%%%%%%%%%%%%%%%%

\newpage
\begin{figure}
\centering
\epsfig{file=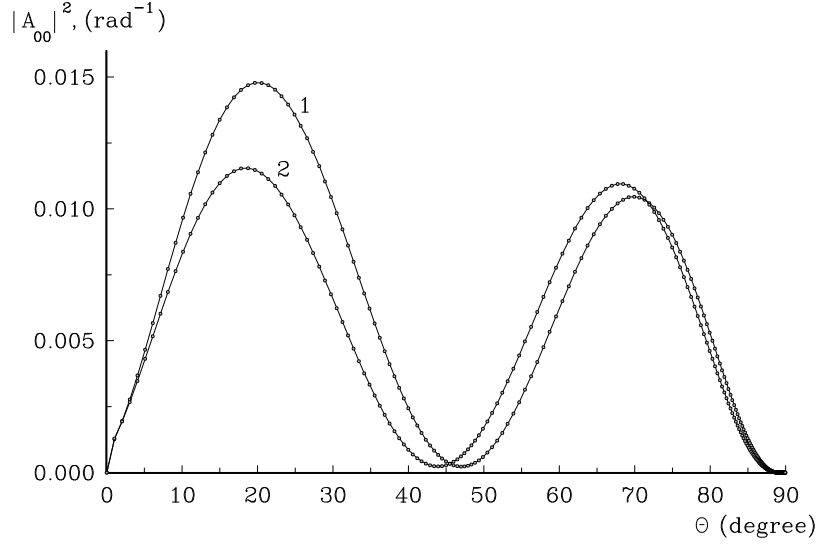,height=9cm}
\caption{The square of the modulus of the Faddeev breakup
amplitude $A_{00}(\theta)$ for HFD-B $^4$He--$^4$He potential at
$E=+1.4$\,mK.  Curve~1 corresponds to the $L=0$, $l=\lambda=0$
partial wave while curve~2 was obtained with the inclusion of
the $L=0$, $l=\lambda=2$ channel. The grid parameters used were
the same as in Table~V.}
\label{Fad-ampl00}
\end{figure}

%%%%%%%%%%%%%%%   FIGURE  3  %%%%%%%%%%%%%%%%%%%%%%%%%%%%%%

\begin{figure}
\centering
\epsfig{file=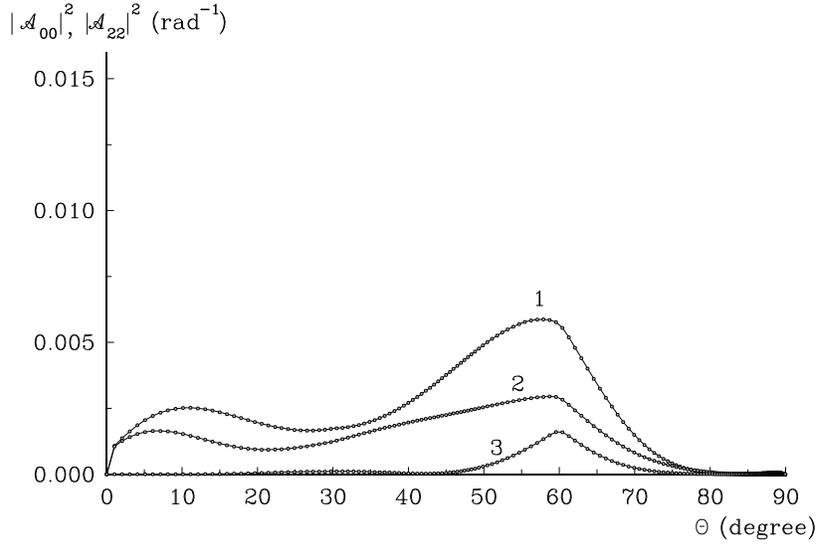,height=9cm}
\caption{The squares of the moduli of the physical breakup
amplitudes ${\cal A}_{00}(\theta)$ (curves~1, 2) and ${\cal
A}_{22}(\theta)$ (curve 3) for the HFD-B $^4$He--$^4$He
potential at $E=+1.4$\,mK.  Curve~1 corresponds to the inclusion
of the $L=0$, $l=\lambda=0$ channel only, while curves~2 and 3
were obtained with the inclusion of both $l=\lambda=0$ and
$l=\lambda=2$ partial waves.  The grid parameters used were the
same as in Table~V.}
\label{fiz-ampl}
\end{figure}

%%%%%%%%%%%%%%%   FIGURE  4  %%%%%%%%%%%%%%%%%%%%%%%%%%%%%%

\begin{figure}
\centering
\psfig{file=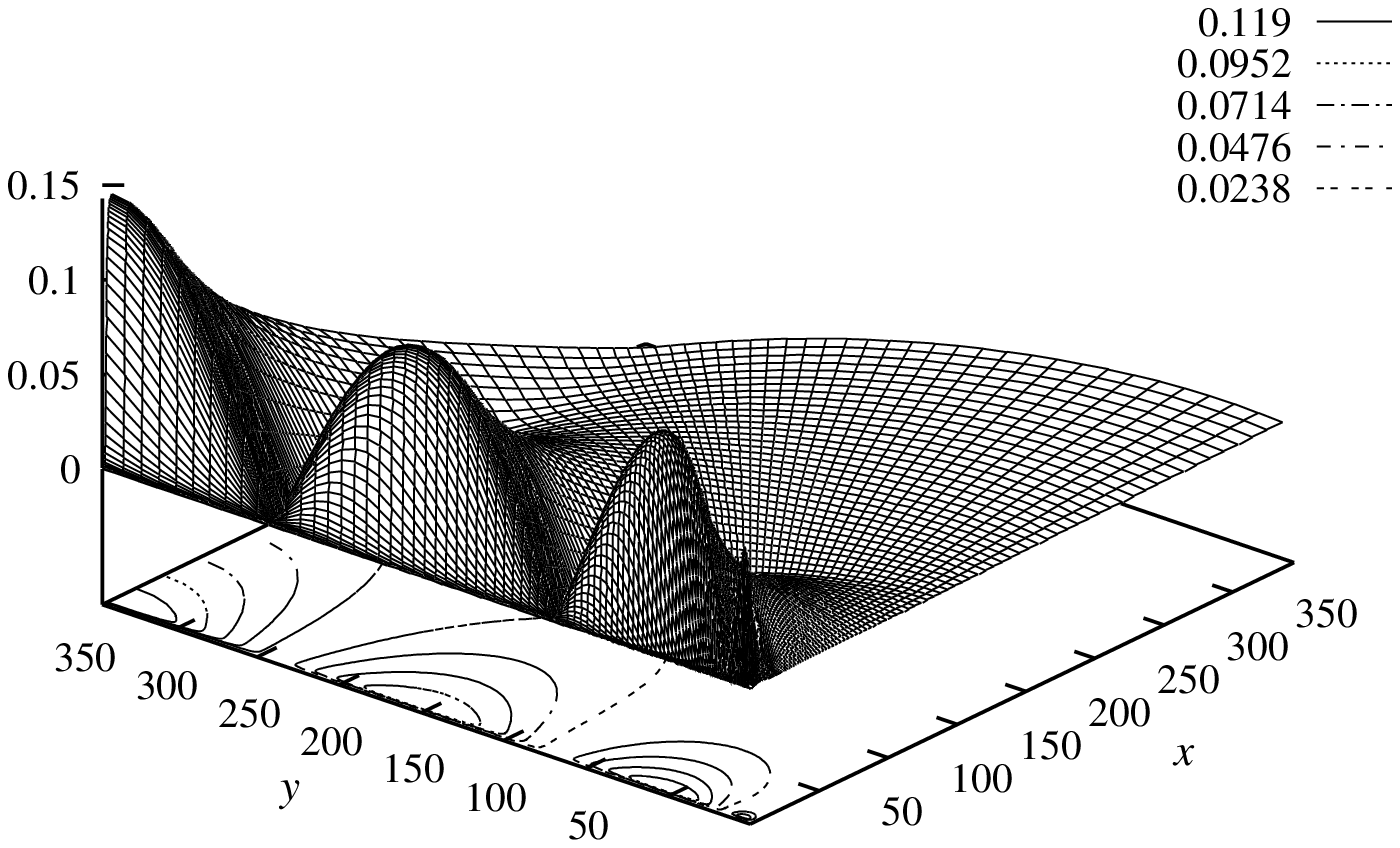,height=9cm}
\caption{Absolute value of the Faddeev component
$\Phi_{000}(x,y,p)$ for the HFD-B $^4$He--$^4$He potential at
$E=+1.4$\,mK.  The grid parameters used were the same as in
Table~V. The values of $x$ and $y$ are in {\AA}.}
\label{fc1a}
\end{figure}

%%%%%%%%%%%%%%%   FIGURE  5  %%%%%%%%%%%%%%%%%%%%%%%%%%%%%%

\begin{figure}
\centering
\psfig{file=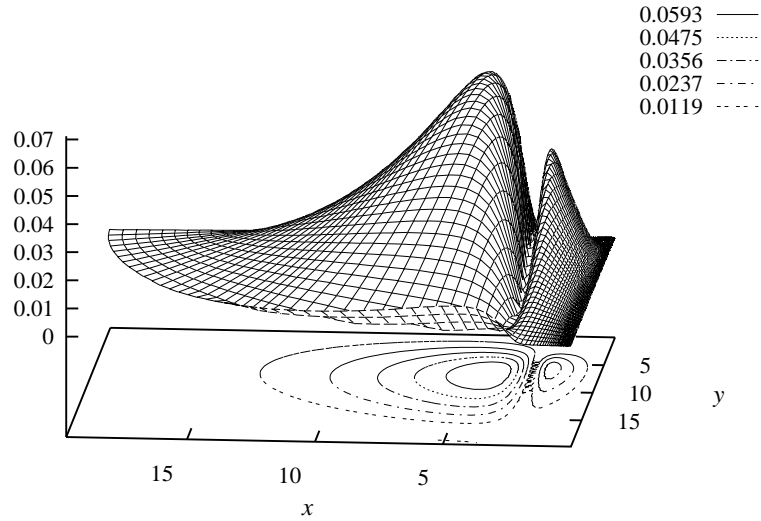,height=9cm}
\caption{Detail of the $|\Phi_{000}(x,y,p)|$ surface
shown in Fig.~4.}
\label{fc1dp}
\end{figure}

%%%%%%%%%%%%%%%   FIGURE  6  %%%%%%%%%%%%%%%%%%%%%%%%%%%%%%

\begin{figure}
\centering
\psfig{file=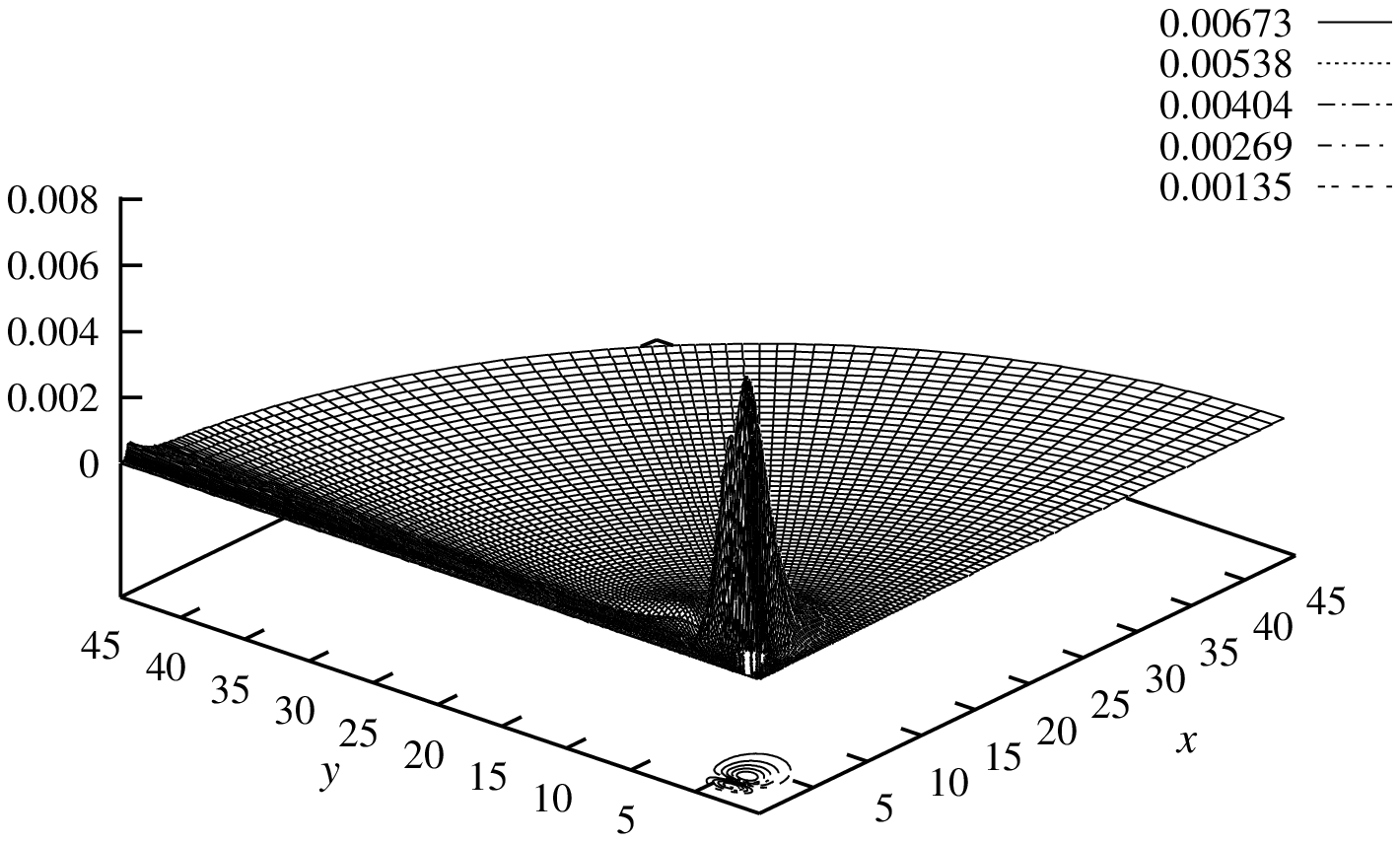,height=9cm}
\caption{Absolute value of the Faddeev component
$\Phi_{220}(x,y,p)$ for the HFD-B $^4$He--$^4$He potential.  at
$E=+1.4$\,mK.  The grid parameters used were the same as in
Table~V. The values of $x$ and $y$ are in {\AA}.}
\label{fc2c}
\end{figure}

%%%%%%%%%%%%%%%   FIGURE  7  %%%%%%%%%%%%%%%%%%%%%%%%%%%%%%

\begin{figure}
\centering
\psfig{file=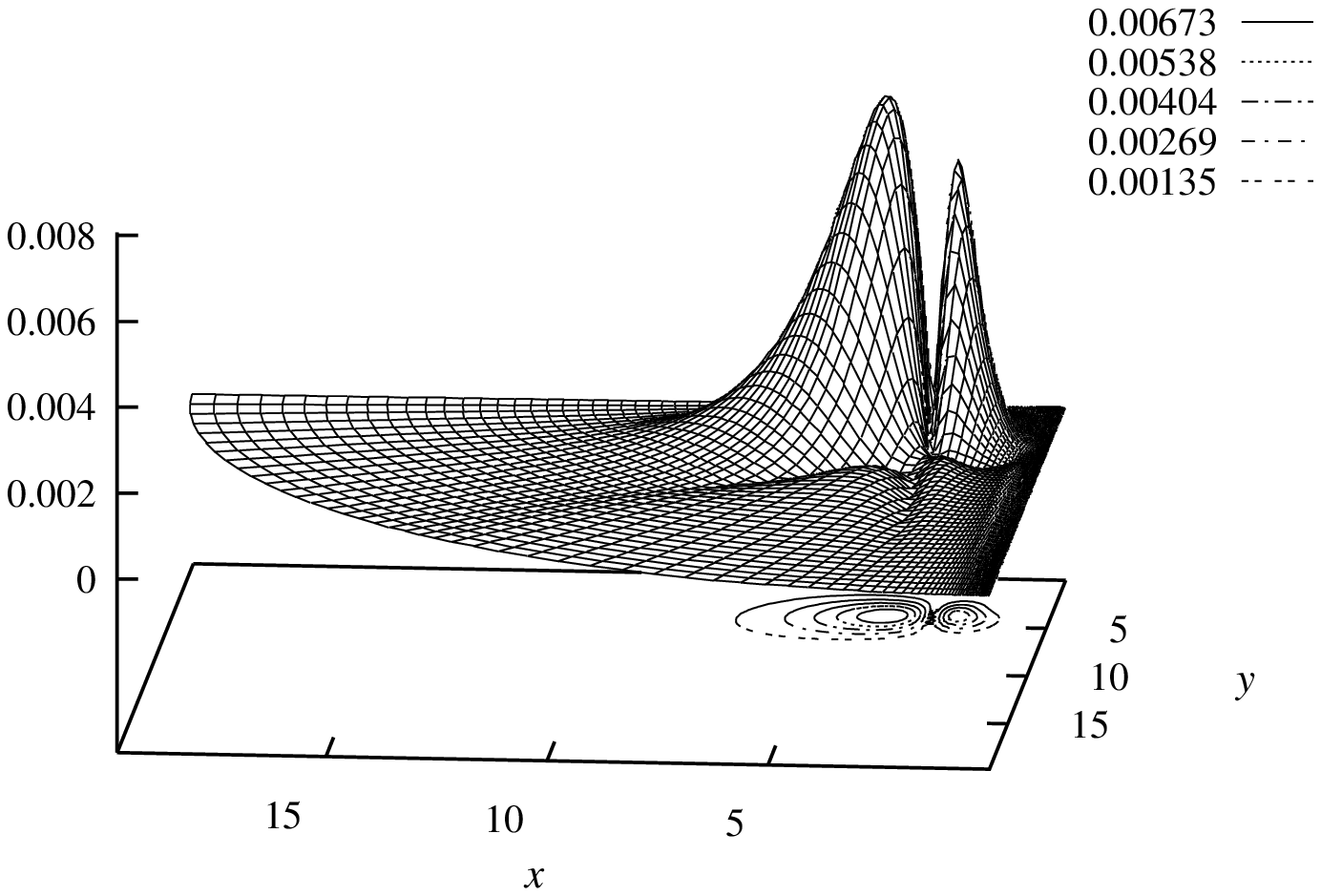,height=9cm}
\caption{Detail of the $|\Phi_{220}(x,y,p)|$ surface
shown in Fig.~6.}
\label{fc2dp}
\end{figure}

%%%%%%%%%%%%%%%%%%%%%%%%%%%%%%%%%%%%%%%%%%%%%%%%%%%%%%%%%%%

\begin{thebibliography}{99}
\bibitem{EfiSch}
       V. N. Efimov, H. Schulz,
     Sov. J. Part. Nucl. {\bf 7}, 349 ( 1976).
\bibitem{MMYa}
       S. P. Merkuriev, A. K. Motovilov,  and S. L. Yakovlev,
         Theor. Math. Phys. {\bf 94},  306 (1993)
       (also see LANL E-print {\tt nucl-th/9606022}).
\bibitem{Faddeev63}
       L. D.  Faddeev,
       {\it  Mathematical
       aspects of the three-body problem in quantum mechanics}
    (Israel Program for Scientific Translations,
       Jerusalem, 1965).
\bibitem{MF}
       L. D.  Faddeev, S. P. Merkuriev,
 {\it Quantum scattering theory for several particle systems}
          (Doderecht: Kluwer Academic Publishers, 1993).
\bibitem{FaddeevMIFI}
      L. D.  Faddeev,
   {\it The integral equation method in scattering theory
    for three and more particles}.
   (Moscow Physics Engineering Institute, Moscow, 1971
    (in Russian)).
\bibitem{KimT}
    Y. E.  Kim,  A. Tubis,
       Phys. Rev. C {\bf 4}, 693 (1971);
	 Phys. Lett. {\bf B 38}, 354 (1972).
\bibitem{Belyaev}
      V. B.  Belyaev,  A. L. Zubarev, Fizika {\bf 3}, 77 (1971).
\bibitem{Brayshaw}
   D. D.  Brayshaw, Phys. Rev. D {\bf 7}, 1835 (1973).
 \bibitem{VEfimov}
       V. Efimov, Yadernaya Fizika (Sov. J.~Nucl. Phys.)
       {\bf 10}, 107 (1969).
\bibitem{KuKhar}
      V. E. Kuzmichev, V. F. Kharchenko,
      Teor. Mat. Fiz. {\bf 31}, 75 (1977).
\bibitem{Potential}
     B.  Schulze, G.  Wildenhain,
    {\it Methoden der  Pothentialtheorie f\"{u}r elliptische differential
     gleihungen beliebiger Ordnung}
  (Academie--Verlag, Berlin, 1977.)
\bibitem{MerMot}
      S. P.  Merkuriev,  A. K. Motovilov,
	Lett. Math. Phys.  {\bf 7}, 497 (1983).
\bibitem{MotovilovVLGU}
       A. K.  Motovilov,
       Vestnik Leningradskogo Universiteta,
       {\bf 22}, 76 (1983).
\bibitem{MM-Kalinin}
     S. P.  Merkuriev,  A. K. Motovilov,
   {\it Theory of Quantum Systems with Strong
   Interaction} (Kalinin University Press, Kalinin, 1983. p. 95--116)
   (Russian).
\bibitem{MotovilovThesis}
       A. K.  Motovilov,
   {\it Three-body quantum problem in the boundary-condition model}
        (PhD thesis (in Russian),  Leningrad State University,
		  Leningrad, 1984).
\bibitem{EChAYa}
       A. A.  Kvitsinsky,  Yu. A. Kuperin,
       S. P.  Merkuriev,  A. K. Motovilov, and S. L. Yakovlev,
    Sov. J. Part. Nucl. {\bf 17}, 113 (1986).
\bibitem{McMillan}
     W. L. McMillan, Phys. Rev. A {\bf 138}, 442 (1983).
\bibitem{Schmid}
     E. W. Schmid, J. Schwager, Y. C. Tang, and R. C. Herndon,
	Physica  {\bf 31}, 1143 (1965).
\bibitem{VM3}
     R. D. Murphy and R. O. Watts, J. Low  Temp. Phys.
     {\bf 2}, 507 (1970).
\bibitem{VM4}
      S. W.  Rick,  D. L. Lynch,  J. D. Doll,
    J. Chem. Phys. {\bf 95}, 3506 (1991).
\bibitem{VMC1}
     K. Schmidt, M. H. Kalos, M. A. Lee, and G. V. Chester,
      Phys. Rev. Lett., {\bf 45}, 573 (1980).
\bibitem{VMC2}
    V. R. Pandharipande, J. G. Zabolitzky, S. C. Pieper, R. B.
    Wiringa, and U. Helmbrecht, Phys. Rev. Lett., {\bf 50}, 1676 (1983).
\bibitem{GF2}
     S. C. Pieper,  R. B. Wiringa, and V. R. Pandharipande,
	   Phys. Rev. B, {\bf 32}, R3341 (1985).
\bibitem{GF3}
    N. Usmani, S. Fantoni, and V. R. Pandharipande,
    Phys. Rev. B, {\bf 26}, 6123 (1983).
\bibitem{GF4}
     S. C. Pieper, in {\em Lecture Notes in Physics}, edited
     by Araki et al., {\bf 198}, page 177, Springer-Verlag, Berlin.
%
\bibitem{Kalos}
   M. H. Kalos, M. A. Lee, P. A. Whitlock , and G. V. Chester,
	 Phys. Rev. B, {\bf 24}, 115(1981).
\bibitem{GF6}
     J. G. Zabolitzky and M. H. Kalos, Nucl. Phys.,
	      {\bf A 356}, 114 (1981).
\bibitem{Nakai}
     S. Nakaichi-Maeda and T. K. Lim, Phys. Rev A, {\bf 28},  692 (1983).
\bibitem{Gloeckle}
       Th. Cornelius, W. Gl\"ockle, J. Chem. Phys.,
	      {\bf 85}, 3906 (1986).
\bibitem{CGM}
      J.  Carbonell,  C. Gignoux,  S. P. Merkuriev,
     Few--Body Systems {\bf 15}, 15 (1993).
\bibitem{Levinger}
      J. S.  Levinger,
     Yadernaya Fizika (Phys. Atom. Nucl.)  {\bf 56}, 106 (1993).
\bibitem{Sof}
      M. Braun, S. A. Sofianos, D. G. Papageorgiou, and I. E. Lagaris,
     Preprint UNISA-NP-96/12 (1996).
\bibitem{EsryLinGreene}
     B.~D.~Esry, C.~D.~Lin, and C.~H.~Greene,
    Phys. Rev.~A {\bf 54}, 394 (1996).
\bibitem{MichaRev81} 
    D. A. Micha, Nucl. Phys. A {\bf 353}, 309 (1981).
\bibitem{Kuppermann} 
    A. Kuppermann, Nucl. Phys. A {\bf 353}, 287 (1981).
%\bibitem{AGS} 
%    E. O. Alt, P. Grassberger, and W. Sandhas,
%    Nucl. Phys. B {\bf 2}, 167 (1967).
\bibitem{KuruogluMicha} Z. C. Kuruoglu, and D. A. Micha, 
    J. Chem. Phys. {\bf 80}, 4262 (1980). 
\bibitem{GhassibChester} H. B. Ghassib, and G. V. Chester, 
    J. Chem. Phys. {\bf 82}, 585 (1984).
\bibitem{March} 
    N. H. March, J. Chem. Phys. {\bf 82}, 587 (1984).
\bibitem{ArgonExp}
    U.  Buck, H. Meyer,
      J. Chem. Phys. {\bf 84}, 4854 (1986).
\bibitem{XenonExp}
     O.  Echt, K.  Sattler, and E. Recknagel,
     Phys. Rev. Lett. {\bf 47}, 1121 (1981).
\bibitem{DimerExp}
     F. Luo,  G. C. McBane, G. Kim,  C. F. Giese,  and  W. R. Gentry,
      J. Chem. Phys. {\bf 98}, 3564 (1993).
\bibitem{Science} W. Sch\"ollkopf and J. P. Toennies,
     Science {\bf 266}, 1345 (1994).
\bibitem{Huber78}
      H. S.  Huber,  T. K. Lim,
	 J. Chem. Phys. {\bf 78}, 1006 (1978).
\bibitem{Aziz79}
     R. A.  Aziz,  V. P. S. Nain,  J. S. Carley,  W. L. Taylor, and
      G. T.  McConville,
     J. Chem. Phys. {\bf 79}, 4330 (1979).
\bibitem{Aziz87}
     R. A. Aziz, F. R. W. McCourt, and C. C. K.  Wong,
      Mol. Phys. {\bf 61}, 1487 (1987).
\bibitem{Aziz91}
       R. A. Aziz and M. J. Slaman,
       J. Chem. Phys. {\bf 94}, 8047 (1991).
\bibitem{Tang95}
      K. T. Tang, J. P. Toennies, and C. L. Yiu,
      Phys. Rev. Lett. {\bf 74}, 1546 (1995).
\bibitem{Efimov} V.~Efimov, Nucl. Phys. A, {\bf 210},
   157 (1973).
\bibitem{MSK-CPL} A. K. Motovilov, S. A. Sofianos, and E. A. Kolganova, 
     Chem. Phys. Lett. {\bf 275}, 168 (1997). 
     LANL E-print {\tt physics/9709037}.
\bibitem{MGL}
     S. P. Merkuriev,  C. Gignoux,  and  A. Laverne,
     Ann. Phys. (N.Y.) {\bf 99}, 30 (1976).
\bibitem{Messiah} A. Messiah. {\it Quantum Mechanics, Vol.~II}
    (North-Holland Publishing Company, Amsterdam, 1966).
\bibitem{Uang}
    Y. H.  Uang, W. C.  Stwalley,
       J. Chem. Phys.  {\bf 76}, 5069 (1982).
\bibitem{yafkm}
     E. A.  Kolganova,  A. K. Motovilov,
     Phys. Atom. Nucl. {\bf 60}, 177 (1997). 
     LANL E-print {\tt nucl-th/9602001} 
     (also see LANL E-print {\tt nucl-th/9702037}).

\end{thebibliography}
\end{document}